%#!tex paper_v2.tex
\input harvmac

\input amssym
\let\includefigures=\iftrue
%
% the following is to use blackboard bold fonts --
%\let\useblackboard=\iftrue
%
% activate this if you don't have them.
%\let\useblackboard=\iffalse
%
% You might also need to remove this line.
\newfam\black
\noblackbox
\includefigures
\message{If you do not have epsf.tex (to include figures),}
\message{change the option at the top of the tex file.}
\def\figin{\epsfcheck\figin}\def\figins{\epsfcheck\figins}
\def\epsfcheck{\ifx\epsfbox\UnDeFiNeD
\message{(NO epsf.tex, FIGURES WILL BE IGNORED)}
\gdef\figin##1{\vskip2in}\gdef\figins##1{\hskip.5in}% blank space instead
\else\message{(FIGURES WILL BE INCLUDED)}%
\gdef\figin##1{##1}\gdef\figins##1{##1}\fi}
\def\DefWarn#1{}

\def\figinsert{\goodbreak\midinsert}
\def\ifig#1#2#3{\DefWarn#1\xdef#1{fig.~\the\figno}
\writedef{#1\leftbracket fig.\noexpand~\the\figno}%
\figinsert\figin{\centerline{#3}}\medskip\centerline{\vbox{\baselineskip12pt
\advance\hsize by -1truein\noindent\footnotefont{\bf
Fig.~\the\figno:} #2}}
\bigskip\endinsert\global\advance\figno by1}
%%%
\else
\def\ifig#1#2#3{\xdef#1{fig.~\the\figno}
\writedef{#1\leftbracket fig.\noexpand~\the\figno}%
%\figinsert\figin{\centerline{#3}}\medskip\centerline{\vbox{\baselineskip12pt
%\advance\hsize by -1truein\noindent\footnotefont{\bf Fig.~\the\figno:} #2}}
%\bigskip\endinsert
\global\advance\figno by1} \fi

%%%%%%% References %%%%%%%

\def\mod{{\rm mod}}
\def\det{{\rm det}}
\def\CP{{\cal P}}
%% MACROS

\def\IL{\relax{\rm I\kern-.18em L}}
\def\IH{\relax{\rm I\kern-.18em H}}
\def\IR{\relax{\rm I\kern-.18em R}}
\def\IC{\relax\hbox{$\inbar\kern-.3em{\rm C}$}}
\def\IZ{\relax\ifmmode\mathchoice
{\hbox{\cmss Z\kern-.4em Z}}{\hbox{\cmss Z\kern-.4em Z}}
{\lower.9pt\hbox{\cmsss Z\kern-.4em Z}} {\lower1.2pt\hbox{\cmsss
Z\kern-.4em Z}}\else{\cmss Z\kern-.4em Z}\fi}
\def\CM {{\cal M}}

\def\CP {{\cal P }}

\def\CV {{\cal V}}
\def\CO {{\cal O}}
\def\CZ {{\cal Z}}

%% MORE MACROS
\def\CM {{\cal M}}

\def\CO {{\cal O}}

\def\CP {{\cal P }}

\def\CV{{\cal V }}
\def\CZ {{\cal Z }}

\def\det{{\rm det}}
\def\Tr{{\rm Tr}}

\font\manual=manfnt \def\dbend{\lower3.5pt\hbox{\manual\char127}}

\def\IZ{\relax\ifmmode\mathchoice
{\hbox{\cmss Z\kern-.4em Z}}{\hbox{\cmss Z\kern-.4em Z}}
{\lower.9pt\hbox{\cmsss Z\kern-.4em Z}} {\lower1.2pt\hbox{\cmsss
Z\kern-.4em Z}}\else{\cmss Z\kern-.4em Z}\fi}

\def\lfm#1{\medskip\noindent\item{#1}}

\def\rt2{\sqrt{2}}
\def\irt2{{1\over\sqrt{2}}}

\def\hat{\widehat}
%  \slashchar puts a slash through a character to represent contraction
%  with Dirac matrices. Use \not instead for negation of relations, and use
%  \hbar for hbar.
\def\slashchar#1{\setbox0=\hbox{$#1$}           % set a box for #1
   \dimen0=\wd0                                 % and get its size
   \setbox1=\hbox{/} \dimen1=\wd1               % get size of /
   \ifdim\dimen0>\dimen1                        % #1 is bigger
      \rlap{\hbox to \dimen0{\hfil/\hfil}}      % so center / in box
      #1                                        % and print #1
   \else                                        % / is bigger
      \rlap{\hbox to \dimen1{\hfil$#1$\hfil}}   % so center #1
      /                                         % and print /
   \fi}

%\DavidTX
\lref\DavidTX{ F.~David, ``Planar Diagrams, Two-Dimensional
Lattice Gravity And Surface Models,'' Nucl.\ Phys.\ B {\bf 257},
45 (1985).
%%CITATION = NUPHA,B257,45;%%
}

%\KazakovDS
\lref\KazakovDS{ V.~A.~Kazakov, ``Bilocal Regularization Of Models
Of Random Surfaces,'' Phys.\ Lett.\ B {\bf 150}, 282 (1985).
%%CITATION = PHLTA,B150,282;%%
}
%\KazakovEA
\lref\KazakovEA{ V.~A.~Kazakov, A.~A.~Migdal and I.~K.~Kostov,
``Critical Properties Of Randomly Triangulated Planar Random
Surfaces,'' Phys.\ Lett.\ B {\bf 157}, 295 (1985).
%%CITATION = PHLTA,B157,295;%%
}
%\AmbjornAZ
\lref\AmbjornAZ{ J.~Ambjorn, B.~Durhuus and J.~Frohlich,
``Diseases Of Triangulated Random Surface Models, And Possible
Cures,'' Nucl.\ Phys.\ B {\bf 257}, 433 (1985).
%%CITATION = NUPHA,B257,433;%%
}
%\DouglasVE
\lref\DouglasVE{ M.~R.~Douglas and S.~H.~Shenker, ``Strings In
Less Than One-Dimension,'' Nucl.\ Phys.\ B {\bf 335}, 635 (1990).
%%CITATION = NUPHA,B335,635;%%
}
%\GrossVS
\lref\GrossVS{ D.~J.~Gross and A.~A.~Migdal, ``Nonperturbative
Two-Dimensional Quantum Gravity,'' Phys.\ Rev.\ Lett.\  {\bf 64},
127 (1990).
%%CITATION = PRLTA,64,127;%%
}
%\BrezinRB
\lref\BrezinRB{ E.~Brezin and V.~A.~Kazakov, ``Exactly Solvable
Field Theories Of Closed Strings,'' Phys.\ Lett.\ B {\bf 236}, 144
(1990).
%%CITATION = PHLTA,B236,144;%%
}
%\DouglasDD
\lref\DouglasDD{ M.~R.~Douglas, ``Strings In Less Than
One-Dimension And The Generalized K-D-V Hierarchies,'' Phys.\
Lett.\ B {\bf 238}, 176 (1990).
%%CITATION = PHLTA,B238,176;%%
}

%\GinspargIS
\lref\GinspargIS{ P.~Ginsparg and G.~W.~Moore, ``Lectures On 2-D
Gravity And 2-D String Theory,'' arXiv:hep-th/9304011.
%%CITATION = HEP-TH 9304011;%%
}

%\DiFrancescoNW
\lref\DiFrancescoNW{ P.~Di Francesco, P.~Ginsparg and
J.~Zinn-Justin, ``2-D Gravity and random matrices,'' Phys.\ Rept.\
{\bf 254}, 1 (1995) [arXiv:hep-th/9306153].
%%CITATION = HEP-TH 9306153;%%
}

%\DornSV
\lref\DornSV{ H.~Dorn and H.~J.~Otto, ``Some conclusions for
noncritical string theory drawn from two and three point functions
in the Liouville sector,'' arXiv:hep-th/9501019.
%%CITATION = HEP-TH 9501019;%%
}
%\ZamolodchikovAA
\lref\ZamolodchikovAA{ A.~B.~Zamolodchikov and
A.~B.~Zamolodchikov, ``Structure constants and conformal bootstrap
in Liouville field theory,'' Nucl.\ Phys.\ B {\bf 477}, 577 (1996)
[arXiv:hep-th/9506136].
%%CITATION = HEP-TH 9506136;%%
}

%\TeschnerYF
\lref\TeschnerYF{ J.~Teschner, ``On the Liouville three point
function,'' Phys.\ Lett.\ B {\bf 363}, 65 (1995)
[arXiv:hep-th/9507109].
%%CITATION = HEP-TH 9507109;%%
}

%\FateevIK
\lref\FateevIK{ V.~Fateev, A.~B.~Zamolodchikov and
A.~B.~Zamolodchikov, ``Boundary Liouville field theory. I:
Boundary state and boundary  two-point function,''
arXiv:hep-th/0001012.
%%CITATION = HEP-TH 0001012;%%
}
%\TeschnerMD
\lref\TeschnerMD{ J.~Teschner, ``Remarks on Liouville theory with
boundary,'' arXiv:hep-th/0009138.
%%CITATION = HEP-TH 0009138;%%
}

%\ZamolodchikovAH
\lref\ZamolodchikovAH{ A.~B.~Zamolodchikov and
A.~B.~Zamolodchikov, ``Liouville field theory on a pseudosphere,''
arXiv:hep-th/0101152.
%%CITATION = HEP-TH 0101152;%%
}
%\PonsotNG
\lref\PonsotNG{ B.~Ponsot and J.~Teschner, ``Boundary Liouville
field theory: Boundary three point function,'' Nucl.\ Phys.\ B
{\bf 622}, 309 (2002) [arXiv:hep-th/0110244].
%%CITATION = HEP-TH 0110244;%%
}
%\McGreevyKB
\lref\McGreevyKB{ J.~McGreevy and H.~Verlinde, ``Strings from
tachyons: The $c = 1$ matrix reloaded,'' arXiv:hep-th/0304224.
%%CITATION = HEP-TH 0304224;%%
}

%\MartinecKA
\lref\MartinecKA{ E.~J.~Martinec, ``The annular report on
non-critical string theory,'' arXiv:hep-th/0305148.
%%CITATION = HEP-TH 0305148;%%
}

%\KlebanovKM
\lref\KlebanovKM{ I.~R.~Klebanov, J.~Maldacena and N.~Seiberg,
``D-brane decay in two-dimensional string theory,'' JHEP {\bf
0307}, 045 (2003) [arXiv:hep-th/0305159].
%%CITATION = HEP-TH 0305159;%%
}
%\McGreevyEP
\lref\McGreevyEP{ J.~McGreevy, J.~Teschner and H.~Verlinde,
``Classical and quantum D-branes in 2D string theory,''
arXiv:hep-th/0305194.
%%CITATION = HEP-TH 0305194;%%
}

%\AlexandrovNN
\lref\AlexandrovNN{ S.~Y.~Alexandrov, V.~A.~Kazakov and
D.~Kutasov, ``Non-perturbative effects in matrix models and
D-branes,'' JHEP {\bf 0309}, 057 (2003) [arXiv:hep-th/0306177].
%%CITATION = HEP-TH 0306177;%%
}

%\KostovUH
\lref\KostovUH{ I.~K.~Kostov, B.~Ponsot and D.~Serban, ``Boundary
Liouville theory and 2D quantum gravity,'' [arXiv:hep-th/0307189].
%%CITATION = HEP-TH 0307189;%%
}

%\DouglasUP
\lref\DouglasUP{ M.~R.~Douglas, I.~R.~Klebanov, D.~Kutasov,
J.~Maldacena, E.~Martinec and N.~Seiberg, ``A new hat for the $c =
1$ matrix model,'' arXiv:hep-th/0307195.
%%CITATION = HEP-TH 0307195;%%
}
%\TakayanagiSM
\lref\TakayanagiSM{ T.~Takayanagi and N.~Toumbas, ``A matrix model
dual of type 0B string theory in two dimensions,'' JHEP {\bf
0307}, 064 (2003) [arXiv:hep-th/0307083].
%%CITATION = HEP-TH 0307083;%%
}

%\KlebanovWG
\lref\KlebanovWG{ I.~R.~Klebanov, J.~Maldacena and N.~Seiberg,
``Unitary and complex matrix models as 1-d type 0 strings,''
arXiv:hep-th/0309168.
%%CITATION = HEP-TH 0309168;%%
}
%\TeschnerQK
\lref\TeschnerQK{ J.~Teschner, ``On boundary perturbations in
Liouville theory and brane dynamics in noncritical string
theories,'' JHEP {\bf 0404}, 023 (2004) [arXiv:hep-th/0308140].
%%CITATION = HEP-TH 0308140;%%
}

%\KarczmarekPV
\lref\KarczmarekPV{ J.~L.~Karczmarek and A.~Strominger, ``Matrix
cosmology,'' JHEP {\bf 0404}, 055 (2004) [arXiv:hep-th/0309138].
%%CITATION = HEP-TH 0309138;%%
}
%\AlexandrovUN
\lref\AlexandrovUN{ S.~Alexandrov, ``$(m,n)$ ZZ branes and the $c
= 1$ matrix model,'' arXiv:hep-th/0310135.
%%CITATION = HEP-TH 0310135;%%
}
%\JohnsonHY
\lref\JohnsonHY{ C.~V.~Johnson, ``Non-perturbative string
equations for type 0A,'' arXiv:hep-th/0311129.
%%CITATION = HEP-TH 0311129;%%
}
%\SeibergNM
\lref\SeibergNM{ N.~Seiberg and D.~Shih, ``Branes, rings and
matrix models in minimal (super)string theory,'' JHEP {\bf 0402},
021 (2004) [arXiv:hep-th/0312170].
%%CITATION = HEP-TH 0312170;%%
}
%\GaiottoYB
\lref\GaiottoYB{ D.~Gaiotto and L.~Rastelli, ``A paradigm of
open/closed duality: Liouville D-branes and the Kontsevich
model,'' [arXiv:hep-th/0312196].
%%CITATION = HEP-TH 0312196;%%
}
%\KostovCY
\lref\KostovCY{ I.~K.~Kostov, ``Boundary ground ring in 2D string
theory,'' [arXiv:hep-th/0312301].
%%CITATION = HEP-TH 0312301;%%
}

%\GrossZZ
\lref\GrossZZ{ D.~J.~Gross and J.~Walcher, ``Non-perturbative RR
potentials in the c(hat) = 1 matrix model,''
[arXiv:hep-th/0312021].
%%CITATION = HEP-TH 0312021;%%
}
%\GukovYP
\lref\GukovYP{ S.~Gukov, T.~Takayanagi and N.~Toumbas, ``Flux
backgrounds in 2D string theory,'' JHEP {\bf 0403}, 017 (2004)
[arXiv:hep-th/0312208].
%%CITATION = HEP-TH 0312208;%%
}
%\AlexandrovKS
\lref\AlexandrovKS{ S.~Alexandrov, ``D-branes and complex curves
in c=1 string theory,'' [arXiv:hep-th/0403116].
%%CITATION = HEP-TH 0403116;%%
}
\lref\KazakovDU{ V.~A.~Kazakov and I.~K.~Kostov, ``Instantons in
non-critical strings from the two-matrix model,''
arXiv:hep-th/0403152.
%%CITATION = HEP-TH 0403152;%%
}
%\DasHW
\lref\DasHW{ S.~R.~Das, J.~L.~Davis, F.~Larsen and
P.~Mukhopadhyay, ``Particle production in matrix cosmology,''
arXiv:hep-th/0403275.
%%CITATION = HEP-TH 0403275;%%
}
\lref\HanadaIM{ M.~Hanada, M.~Hayakawa, N.~Ishibashi, H.~Kawai,
T.~Kuroki, Y.~Matsuo and T.~Tada, ``Loops versus Matrices -The
nonperturbative aspects of noncritical string-,''
[arXiv:hep-th/0405076].
%%CITATION = HEP-TH 0405076;%%
}
%\MooreIR
\lref\MooreIR{ G.~W.~Moore, N.~Seiberg and M.~Staudacher, ``From
loops to states in 2-D quantum gravity,'' Nucl.\ Phys.\ B {\bf
362}, 665 (1991).
%%CITATION = NUPHA,B362,665;%%
}
%\MooreAG
\lref\MooreAG{ G.~W.~Moore and N.~Seiberg, ``From loops to fields
in 2-D quantum gravity,'' Int.\ J.\ Mod.\ Phys.\ A {\bf 7}, 2601
(1992).
%%CITATION = IMPAE,A7,2601;%%
}
%\SeibergEB
\lref\SeibergEB{ N.~Seiberg, ``Notes On Quantum Liouville Theory
And Quantum Gravity,'' Prog.\ Theor.\ Phys.\ Suppl.\  {\bf 102},
319 (1990).
%%CITATION = PTPSA,102,319;%%
}
%\PolchinskiMH
\lref\PolchinskiMH{ J.~Polchinski, ``Remarks On The Liouville
Field Theory,'' UTTG-19-90
%\href{http://www.slac.stanford.edu/spires/find/hep/www?r=uttg-19-90}{SPIRES
%entry}
{\it Presented at Strings '90 Conf., College Station, TX, Mar
12-17, 1990}}

%\DaulBG
\lref\DaulBG{ J.~M.~Daul, V.~A.~Kazakov and I.~K.~Kostov,
``Rational theories of 2-D gravity from the two matrix model,''
Nucl.\ Phys.\ B {\bf 409}, 311 (1993) [arXiv:hep-th/9303093].
%%CITATION = HEP-TH 9303093;%%
}

%\AmbjornJI
\lref\AmbjornJI{ J.~Ambjorn, J.~Jurkiewicz and Y.~M.~Makeenko,
``Multiloop Correlators For Two-Dimensional Quantum Gravity,''
Phys.\ Lett.\ B {\bf 251}, 517 (1990).
%%CITATION = PHLTA,B251,517;%%
}
%\KostovCG
\lref\KostovCG{ I.~K.~Kostov, ``Strings with discrete target
space,'' Nucl.\ Phys.\ B {\bf 376}, 539 (1992)
[arXiv:hep-th/9112059].
%%CITATION = HEP-TH 9112059;%%
}
%\HosomichiXC
\lref\HosomichiXC{ K.~Hosomichi, ``Bulk-boundary propagator in
Liouville theory on a disc,'' JHEP {\bf 0111}, 044 (2001)
[arXiv:hep-th/0108093].
%%CITATION = HEP-TH 0108093;%%
}
%\PonsotSS
\lref\PonsotSS{ B.~Ponsot, ``Liouville theory on the pseudosphere:
Bulk-boundary structure constant,'' arXiv:hep-th/0309211.
%%CITATION = HEP-TH 0309211;%%
}
%\MartinecHT
\lref\MartinecHT{ E.~J.~Martinec, G.~W.~Moore and N.~Seiberg,
``Boundary operators in 2-D gravity,'' Phys.\ Lett.\ B {\bf 263},
190 (1991).
%%CITATION = PHLTA,B263,190;%%
}
%\LianGK
\lref\LianGK{ B.~H.~Lian and G.~J.~Zuckerman, ``New Selection
Rules And Physical States In 2-D Gravity: Conformal Gauge,''
Phys.\ Lett.\ B {\bf 254}, 417 (1991).
%%CITATION = PHLTA,B254,417;%%
}
\lref\PasquierJC{ V.~Pasquier, ``Two-Dimensional Critical Systems
Labelled By Dynkin Diagrams,'' Nucl.\ Phys.\ B {\bf 285}, 162
(1987).
%%CITATION = NUPHA,B285,162;%%
}
%\DavidNQ
\lref\DavidNQ{ J.~R.~David, S.~Minwalla and C.~Nunez, ``Fermions
in bosonic string theories,'' JHEP {\bf 0109}, 001 (2001)
[arXiv:hep-th/0107165].
%%CITATION = HEP-TH 0107165;%%
}

%\draftmode

\newbox\tmpbox\setbox\tmpbox\hbox{\abstractfont }
\Title{\vbox{\baselineskip12pt\hbox{EFI-04-19} \hbox{PUPT-2121}}}
{\vbox{\centerline{Annulus Amplitudes and ZZ Branes in}\smallskip
\centerline{ Minimal String Theory
}}}
\smallskip
\centerline{ David Kutasov,$^1$ Kazumi Okuyama,$^1$ Jongwon
Park,$^1$  Nathan Seiberg$^2$ and David Shih$^3$}
\smallskip
\bigskip
\centerline{$^1${\it Enrico Fermi Institute, University of
Chicago, Chicago, IL 60637, USA}}
\medskip
\centerline{$^2${\it School of Natural Sciences, Institute for
Advanced Study, Princeton, NJ 08540 USA}}
\medskip
\centerline{$^3${\it Department of Physics, Princeton University,
Princeton, NJ 08544 USA}}
\bigskip
\vskip 1cm

\noindent We study the annulus amplitudes of $(p,q)$ minimal
string theory. Focusing on the ZZ-FZZT annulus amplitude as a
target-space probe of the ZZ brane, we use it to confirm that the
ZZ branes are localized in the strong-coupling region. Along the
way we learn that the ZZ-FZZT open strings are fermions, even
though our theory is bosonic! We also provide a geometrical
interpretation of the annulus amplitudes in terms of the Riemann
surface $\CM_{p,q}$ that emerges from the FZZT branes. The ZZ-FZZT
annulus amplitude measures the deformation of $\CM_{p,q}$ due to
the presence of background ZZ branes; each kind of ZZ-brane
deforms only one $A$-period of the surface. Finally, we use the
annulus amplitudes to argue that the ZZ branes can be regarded as
``wrong-branch" tachyons which violate the bound $\alpha <Q/2$.

\vskip 1cm

\Date{June, 2004}
\vfil\eject

\newsec{Introduction}

Minimal string theories are string theories based on the $(p,q)$
minimal conformal field theories coupled to Liouville. They were
first solved using their description in terms of matrix models
\refs{\DavidTX\KazakovDS\KazakovEA\AmbjornAZ\DouglasVE\GrossVS
\BrezinRB-\DouglasDD} (for reviews, see e.g.
\refs{\GinspargIS,\DiFrancescoNW}).  This tractable
description allows explicit calculations of many physical
quantities of interest.  In particular, it has already offered
many insights regarding nonperturbative phenomena, D-branes and
holography.

Recent progress in the study of Liouville theory
\refs{\DornSV\ZamolodchikovAA-\TeschnerYF} and its D-branes
\refs{\FateevIK\TeschnerMD\ZamolodchikovAH-\PonsotNG} has
motivated the renewed interest in the subject
\refs{\McGreevyKB\MartinecKA\KlebanovKM\McGreevyEP\AlexandrovNN
\KostovUH\DouglasUP\TakayanagiSM\KlebanovWG\TeschnerQK
\KarczmarekPV\AlexandrovUN\AlexandrovNN\KlebanovWG\JohnsonHY
\SeibergNM\GaiottoYB\GrossZZ\GukovYP\KostovCY\AlexandrovKS
\KazakovDU\DasHW-\HanadaIM}. In this note we continue the investigation
started in \SeibergNM, whose purpose was to explore the D-branes
of minimal string theory and to derive the matrix model starting
from the worldsheet description. Following \MartinecKA\ we will
focus on the annulus diagram as an interesting diagnostic of the
theory. We will use it to understand the connection between the
target space fields in the theory, the D-branes and their
interpretation in terms of an auxiliary Riemann surface.

Before we begin, let us briefly review $(p,q)$ minimal string
theory. These theories consist of $(p,q)$ minimal CFT coupled to
Liouville theory with background charge $Q=b+b^{-1}$, and they
exist for all relatively prime integers $p$ and $q$. Our
convention throughout will be $p<q$. The parameter $b$, entering
also in the Liouville potential $\mu e^{2b\phi}$, should be set to
\eqn\introb{ b=\sqrt{p\over q} } in order for the total central
charge of the Liouville and the minimal model to be $26$. In the
following, we will set $\mu=1$ for simplicity (after rescaling it
by $\pi\gamma(b^2)$, as in \SeibergNM).

Although the minimal string theories have physical closed-string
vertex operators at every ghost number less than or equal to one,
in this paper we will focus our attention on the ``tachyon"
operators $\CT_{r,s}$ at ghost number one and the ground ring
operators $\hat\CO_{r,s}$ at ghost number zero. Both sets of
operators are labelled by integers $r=1,\dots,p-1$ and
$s=1,\dots,q-1$, although the tachyons obey an additional
reflection relation $\CT_{p-r,q-s}=\CT_{r,s}$ that cuts their
number in half.

One of the results of \SeibergNM\ was the emergence of an
auxiliary Riemann surface $\CM_{p,q}$ which underlies many, if not
all, of the features of minimal string theory. The Riemann surface
appeared from the study of a class of D-branes of minimal string
theory called FZZT branes \refs{\FateevIK,\TeschnerMD}. It is
described by the algebraic equation \eqn\introMpq{ F(x,y)\equiv
T_q(x)-T_p(y)=0 } where $T_p(\cos\theta)=\cos p\theta$ is the
Chebyshev polynomial of the first kind. Here
 \eqn\introxy{x=\mu_B,\qquad y=\partial_{\mu_B}Z(\mu_B) }
correspond to the boundary cosmological constant and the
derivative of the disk amplitude of the FZZT brane, respectively.
It is often convenient to label the FZZT branes by the auxiliary
parameter $\sigma$, in terms of which
 \eqn\introsigma{ x=\cosh\pi b\sigma,\qquad y=\cosh\pi
 b^{-1}\sigma ~.}
Viewed as a branched cover of the complex $x$-plane, $\CM_{p,q}$
clearly has $p$ different sheets. The surface can be uniformized
(i.e.\ reduced to a single copy of the complex plane) by
introducing the parameter
 \eqn\introunif{ z=\cosh{\pi \sigma\over\sqrt{pq}} }
in terms of which $x=T_p(z)$ and $y=T_q(z)$.

Perhaps the most important feature of $\CM_{p,q}$ is that it has
singularities where \eqn\introsingdef{ F=\partial_x F=\partial_y
F=0\ . } The singularities are labelled by integers $m=1,\dots,
p-1$, $n=1,\dots,q-1$ satisfying $qm-pn>0$, and they are located
at \eqn\introsing{ x_{mn}=(-1)^m\cos{\pi n p\over q},\qquad
y_{mn}=(-1)^n\cos{\pi m q\over p} ~.} In terms of the uniformizing
parameter, the singularities are points on $\CM_{p,q}$ which
correspond to two different values of $z$: \eqn\introzmn{
z_{mn}^{\pm}=\cos{\pi(mq\pm np)\over pq}\ . } The singularities
can also be thought of as pinched cycles of a higher-genus
surface. Correspondingly, we can define a canonical basis of $A$
and $B$-cycles on $\CM_{p,q}$ labelled by $m$ and $n$, with
$A_{mn}$ ($B_{mn}$) circling around (passing through) the $(m,n)$
singularity. It turns out that these cycles are closely related to
another class of D-branes, the ZZ branes of minimal string theory
 \refs{\ZamolodchikovAH}.
The independent ZZ branes are also labelled by integers $(m,n)$
and are in one-to-one correspondence with the singularities of
$\CM_{p,q}$.

In section 2, we start from the worldsheet description and derive
simple formulas for the annulus amplitude between two FZZT branes
 \eqn\introann{ Z(\sigma|\sigma') = \log\left({z-z'\over
 T_p(z)-T_p(z') }\right) ~,}
between an $(m,n)$ ZZ brane and an FZZT brane
\eqn\introannII{
Z(m,n|\sigma)=\log\left({z-z_{m,n}^-\over z-z_{m,n}^+}\right) ~,
}
and between two ZZ branes labelled by $(m,n)$ and $(m',n')$
\eqn\introannIII{
Z(m,n|m',n')= \log {(z^+_{m,n} -
 z^+_{m^\prime,n^\prime})(z^-_{m,n} -
 z^-_{m^\prime,n^\prime}) \over (z^+_{m,n} -
 z^-_{m^\prime,n^\prime})(z^-_{m,n} - z^+_{m^\prime,n^\prime}) }~.
 }
Motivated by \refs{\MooreIR,\MooreAG}, we take the inverse Laplace
transform of $Z(m,n|\sigma)$ with respect to the boundary
cosmological constant to obtain the ``minisuperspace amplitude"
$\Psi_{m,n}(\ell)$ at fixed loop length $\ell$. This quantity is a
target-space probe of the background $(m,n)$ ZZ brane, and we use
the large and small $\ell$ behavior of $\Psi_{m,n}(\ell)$ to
verify that the ZZ branes are located deep in the strong-coupling
region $\phi\to+\infty$ of the Liouville direction.

In section 3, we extend the analysis of the deformations of
$\CM_{p,q}$ given in \SeibergNM, at the same time providing a
geometric interpretation for the ZZ-FZZT annulus amplitude. There
are two types of deformations of $\CM_{p,q}$:
singularity-preserving deformations, which change only the
$B$-cycles of $\CM_{p,q}$; and singularity-destroying
deformations, which change both the $A$ and $B$-cycles. In
\SeibergNM, the former class of deformations was considered in
detail, and it was shown that they correspond to the addition of
local closed-string operators to the worldsheet action. Here we
study the space of singularity-destroying deformations, and we
argue that this space is spanned by the ZZ brane deformations. The
main point is that adding $(m,n)$ ZZ branes to the background
deforms the FZZT disk amplitude to leading order by
$Z(m,n|\sigma)$, so that the deformation of the curve \introMpq\
is
\eqn\introdefZZann{
\delta_{m,n} F \sim U_{p-1}(y)\partial_x Z(m,n|\sigma)
}
where $U_{p-1}(\cos\theta)={\sin(p\theta )\over \sin(\theta)}$ is
a Chebyshev polynomial of the second kind. Starting from
\introannII\ and \introdefZZann, we show that the deformation can
be written as a polynomial in $x$ and $y$
 \eqn\introdefZZ{\delta_{m,n}F \sim \sum_{r,s} U_{s-1}(x_{mn})
 U_{r-1}(y_{mn})U_{s-1}(x)U_{p-r-1}(y) ~,}
and we demonstrate that it deforms only the $(m,n)$ $A$-cycle of
the surface, leaving the other $A$-cycles unchanged. Thus, the
effect on $\CM_{p,q}$ of inserting background $(m,n)$ ZZ branes is
to open up the $(m,n)$ singularity into a regular cycle. In this
sense, the ZZ branes ``diagonalize" the $A$-cycle deformations of
$\CM_{p,q}$.

This raises the question of what closed-string
operators diagonalize the $B$-cycle deformations of $\CM_{p,q}$.
Focusing on the tachyons $\CT_{r,s}$, we find the linear
combinations $\tilde\CT_{m,n}$ that deform only the $(m,n)$
$B$-cycle of the surface:
\eqn\introTmn{ \tilde\CT_{m,n} = \sum_{r,s}
U_{s-1}(x_{mn})U_{r-1}(y_{mn}) \CT_{r,s} ~.} We show that these
correspond to eigenstates of the ground ring, and that their
eigenvalues are precisely the singularities of $\CM_{p,q}$.

The similarity between the deformations \introdefZZ\ and
\introTmn\ hints at a duality between the ZZ branes and the
tachyon eigenstates. In section 4, we attempt to make this duality
more precise. We propose that the ZZ branes can be thought of as
``wrong-branch" tachyons, i.e.\ tachyons which violate the usual
bound $\alpha<Q/2$ \SeibergEB. Our proposal is motivated by the
following target-space intuition. In the first quantized
description of the theory, the background at large negative $\phi$
(weak coupling) is determined by solving the Wheeler-deWitt
equation \refs{\SeibergEB,\PolchinskiMH,\MooreIR}. This is a
second-order differential equation in $\phi$, and as such, it has
two linearly independent solutions.  The solution which decays in
the strong coupling end $\phi \to +\infty$ corresponds to closed
string states which satisfy the bound $\alpha<Q/2$. It describes
the deformation of the system by closed strings at the weak
coupling end $\phi \to -\infty$ where it diverges. It is then
natural to ask: what does the other solution correspond to? It
satisfies the field equations but does not respect the boundary
conditions; it can be taken to decay at the weak coupling end
$\phi \to -\infty$ but it diverges at the strong coupling end. As
a simple example of a solution which has a singularity, consider
the $1\over r^2$ solution of the electric field in $3+1 $
dimensions. The singularity at $r=0$ means that a charged particle
is present there.  However, in general, the existence of such a
solution cannot be used as an argument for the existence of
charged particles.  In our case, as in other similar situations in
string theory, this background turns out to be sourced by a brane.
This is the ZZ-brane.  Indeed, the number of distinct ZZ-branes is
the same as the number of tachyons in the system \SeibergNM.  As
we will show, appropriate linear combinations of these ZZ-branes
source the different tachyons which violate the bound $\alpha<Q/2$
and have the ``wrong'' $\phi$ dependence.

Finally, in section 5, we consider in detail the models with
$(p,q)=(2,2k-1)$ and show how many aspects of the general analysis
simplify for these models. In appendix A, we study more general
annulus amplitudes, and we use these to test the FZZT
identification formulas of \SeibergNM. Appendix B contains a
discussion of the proper normalization of the ZZ boundary states
in minimal string theory. We argue that the boundary states should
be normalized with an extra minus sign relative to the
normalization in Liouville theory alone.  This sign leads to the
surprising result that the open strings stretched between ZZ and
FZZT branes are fermions. (The emergence of fermions in the
bosonic string has already been discussed in \DavidNQ.)

\newsec{Annulus Amplitudes}

\subsec{FZZT-FZZT annulus amplitude}

In this section, we will study the annulus amplitude between two
FZZT branes, using the continuum Liouville approach. This was
computed in \MartinecKA; we quote here the result:
\eqn\annulusbare{
Z(\sigma|\sigma') = {1\over
p}\sum_{j=1}^{p-1}\int_{-\infty}^{\infty} d\nu {\sin^2 (\pi j /
p)\cos (\pi\sigma\nu) \cos (\pi\sigma'\nu) \over \nu
\sinh(\pi\nu/b)\Big(\cosh(\pi\nu/b)-\cos(\pi j/p)\Big)}\ .
}
Here we have implicitly chosen the matter boundary state to be the
$(1,1)$ Cardy state (or $a=c=1$ in the notation of \MartinecKA).
This choice of matter state will be assumed throughout, and to
simplify the notation, we will continue to suppress the matter
label. We lose no generality in doing so, since FZZT branes with
other matter states can be written as a linear combination of
branes with matter state $(1,1)$, up to BRST exact terms
\SeibergNM. (We will discuss the FZZT identifications further in
appendix A.) Note that the integral \annulusbare\ has a divergence
due to the double pole at $\nu=0$; we will regularize this by
replacing the factor $1/\nu$ by $\nu/(\nu^2+\varepsilon^2)$
\eqn\annulusj{
Z(\sigma|\sigma') = {1\over
p}\sum_{j=1}^{p-1}\int_{-\infty}^{\infty} d\nu {\nu\over\nu^2+\varepsilon^2}
{\sin^2 (\pi j/ p)\cos (\pi\sigma\nu) \cos (\pi\sigma'\nu) \over
\sinh(\pi\nu/b)\Big(\cosh(\pi\nu/b)-\cos(\pi j/p)\Big)}\ .
}
In appendix A, the summation over $j$ is performed, leading to
\eqn\annulus{
Z(\sigma|\sigma') =
\int_{-\infty}^{\infty} d\nu {\nu\over\nu^2+\varepsilon^2}
{\cos (\pi\sigma\nu) \cos (\pi\sigma'\nu)\sinh(\pi(p-1)\nu/b) \over
\sinh(\pi\nu/b)\sinh(\pi p\nu/b)}\ .
}
Now let us proceed to evaluate this integral in two steps.

The first step is to close the contour of integration either into
the upper or the lower half-plane, so as to pick up the poles at
\eqn\poles{
\nu=\pm i\varepsilon\quad\hbox{and}\quad \nu=
{ik\over\sqrt{pq}}\quad \hbox{for $k\in {\Bbb Z}$}\quad(k\not=0)~.
}
Our task is slightly complicated by the fact that the particular
contour deformation we must use depends on the values of $\sigma$
and $\sigma'$. For instance, if $\sigma$ and $\sigma'$ are real
and $\sigma>\sigma'>0$, then we can write
\eqn\split{
\cos \pi\sigma\nu = {e^{i\pi\sigma\nu}+e^{-i\pi\sigma\nu}\over 2}
}
in the integrand of \annulus, in which case the integral over the
first (second) exponential can be deformed to pick up the poles in
the upper (lower) half-plane. The poles \poles\ with $k=0\,\mod\
p$ and $k\ne0\,\mod\ p$ behave rather differently, leading to two
separate sums. Including the pole at $\nu=\pm i\varepsilon$, we
obtain
\eqn\annulussimp{\eqalign{
&Z(\sigma|\sigma') = {p-1\over p}\sum_{k=1}^{\infty}
 {2\over  k} e^{-{\pi p k\sigma\over\sqrt{pq}}} \cosh {\pi
 p k\sigma'\over\sqrt{pq}}-\sum_{k=1\atop k\ne 0\,\mod\,p}^{\infty}
{2\over k}e^{-{\pi k\sigma\over\sqrt{pq}}} \cosh {\pi
k\sigma'\over\sqrt{pq}} \cr
 &\qquad\qquad\qquad\qquad+ \left({p-1\over\varepsilon\sqrt{pq}}-(p-1){\pi\sigma\over\sqrt{pq}} +
    {\cal O}(\varepsilon) \right)\qquad (\sigma>\sigma'>0)~.
}}

The second step in evaluating \annulus\ is to drop the divergent
piece and evaluate the sums in \annulussimp. This yields a simple,
closed-form expression for the annulus amplitude
\eqn\annulussimpIII{
 Z(\sigma|\sigma') = \log\left(\cosh{\pi
\sigma\over\sqrt{pq}}-\cosh{\pi\sigma'\over\sqrt{pq}}\over
{\cosh{p\pi \sigma\over\sqrt{pq}}-\cosh{p\pi\sigma'\over\sqrt{pq}}
}\right)~.
 }
The analytic continuation of this expression to all $\sigma$,
$\sigma'$ is self-evident. An important check of this formula is
that it agrees (up to an overall sign) with the
two-macroscopic-loop amplitude obtained in the two-matrix model
\DaulBG\ which generalizes expressions that had been found earlier
in special cases \refs{\MooreIR\MooreAG\AmbjornJI-\KostovCG}.
 Moreover, one can easily see that
\annulussimpIII\ depends only on the ``uniformizing parameter" $z$
defined in \introunif. In terms of $z$, \annulussimpIII\ becomes
simply
 \eqn\annulusfin{ Z(\sigma|\sigma') =\log\left(z-z' \over T_p(z)-
 T_p(z')\right)=\log\left(z-z' \over x- x'\right) ~.}

Evidently, the annulus amplitude has some rather interesting
properties when viewed as a function of $z$ and $z'$. It is finite
when $z=z'$ for generic $z$, but it diverges when $T_p(z)=T_p(z')$
with $z\ne z'$. It also diverges when $z=z'$ and $U_{p-1}(z)=0$,
which corresponds to the ends of the cuts in the various sheets.
In other words, if we think of the Riemann surface as a
$p$-sheeted cover of the complex $x=T_p(z)$ plane parametrized by
$z$, the annulus amplitude is {\it finite} when $x\to x'$ on the
same sheet (except at the ends of the cuts), but {\it diverges}
when $x\to x'$ on different sheets.

The expression \annulusfin\ is closely related to the partition
function of two FZZT branes.  In the matrix model, the FZZT brane
is identified with the macroscopic loop operator
$W(x)=\Tr\log(x-M)$. Thus it is represented by an insertion of
\eqn\loopinsert{
e^{W(x)} =\det (x-M)
}
into the matrix integral. Consider now the two point function
$\langle\det (x-M)\det (x'-M)\rangle$. It should be separated into
connected and disconnected contributions
 \eqn\twodb{\langle\det (x-M)\det (x'-M)\rangle = \langle\det
 (x-M)\rangle \langle \det (x'-M)\rangle e^{Z(x,x')} }
where the exponent $Z(x,x')$ is given by a sum of connected
diagrams
 \eqn\Zcon{Z(x,x') = \sum_{m,n\ge 1} {1 \over m! n!}
 \langle W(x)^m W(x')^n\rangle_c ~.}
These connected correlation functions are computed in string
theory by worldsheets having $m$ boundaries with $\mu_B=x$ and $n$
boundaries with $\mu_B=x'$.  To leading order in $1/N$, $Z(x,x')$
is just given by the annulus amplitude \annulusfin, which leads to
 \eqn\twodba{\langle\det (x-M)\det (x'-M)\rangle = \langle\det
 (x-M)\rangle \langle \det (x'-M)\rangle\left({z-z' \over x-
 x'} +\CO\Big( {1\over N}\Big)\right)~.}
Clearly the operator $\det(x-M)$ is a boson, and the two point
function \twodba\ is invariant under the interchange of $x$ and
$x'$. However, comparison with the matrix model suggests that
$\det(x-M)$ can be interpreted as creating a {\it fermionic} probe
eigenvalue. This suggests that one should combine $\det(x-M)$ with
some kind of cocycle factor to produce a fermionic operator. This
might also have the effect of removing the denominator $x-x'$ of
\twodba\ (see also the discussion at the end of this subsection).

Finally, let us comment briefly on the form of the sum in
\annulussimp. We can view this as a sum over physical states with
$k\ne 0\,\mod\,p$ and unphysical states with $k=0\,\mod\,p$. To
understand how the latter arise, it is useful to consider the annulus
amplitude for fixed boundary length $\ell$ (i.e.\ the inverse Laplace
transform of the annulus with respect to $\mu_B$ and $\mu_B'$).
According to \MartinecKA, this has the form
 \eqn\annfixedl{
 Z(\ell_1,\ell_2) \sim \int_{-\infty}^{\infty}d\nu\,
 G(\nu)\psi_{\nu}(\ell_1)\psi_{\nu}(\ell_2)
 }
where
 \eqn\psifixedl{
 \psi_{\nu}(\ell) = K_{2i\nu/b}(\ell)\sqrt{\nu\sinh (2\pi \nu/b)}
 }
are the normalized wavefunctions and
 \eqn\propfixedl{ G(\nu) = \sum_{j=1}^{p-1}{\sin^2 (\pi j/ p)\over \cosh(2\pi\nu/b)-\cos(\pi j/p)} }
is the propagator. Notice how it only has poles at the locations
of the physical states $k\ne 0\,\mod\,p$. This means that when we
close the contour to pick up these poles, we will arrive at a sum
over only the physical states
 \eqn\annfixedlsum{ Z(\ell_1,\ell_2)
\sim \sum_{k=1\atop k\ne 0\,\mod\,p}^{\infty} k \sin (\pi k/ p)
K_{k\over p}(\ell_1)I_{k\over
 p}(\ell_2) \qquad (\ell_1> \ell_2) ~.}
This form of the $\ell$-space annulus was obtained previously for
various special cases in \MooreAG. This expression \annfixedlsum\
is similar to \annulussimp, but here the unphysical states with
$k=0\,\mod\,p$ do not contribute. It is only in the Laplace
transform to fixed $\sigma$ and $\sigma'$ that we obtain the
unphysical poles in the propagator. In fact, it is not difficult
to see that these poles arise from the $\ell_1$, $\ell_2\to 0$
part of the integral. Thus we can think of these poles as the
contribution to the annulus amplitude from zero-area worldsheets.

It is well-known that zero-area surfaces contribute non-universal
terms to the path integral that are analytic in $\mu$ \SeibergEB.
To see this explicitly, we can restore factors of $\mu$ (which was
set to one) and express the amplitude as a function of $\mu$ and
$\mu_B$. From \annulussimp--\annulusfin, we see that the
unphysical poles sum up to produce the $\log(x-x') =
\log(\mu_B-\mu_B') - {1\over 2} \log \mu$ term (up to a
contribution proportional to $\sigma$) in the annulus amplitude.
The $\log\mu$ piece is unimportant for our present discussion.
What matters is that the rest is independent of $\mu$. This
suggests that we should focus our attention on just the
$\log(z-z')$ part of \annulusfin. (Equivalently, we should
differentiate \annulusfin\ with respect to $\mu$ and focus on the
features of the resulting function.) Then
$Z(\sigma|\sigma')=\log(z-z')+\dots$ is relatively simple -- it
diverges when the two FZZT branes collide at the same point in
$\CM_{p,q}$.

We conclude that up to non-universal terms the amplitude is
$\log(z-z')$. This discussion also explains the comment above that
the denominator $x-x'$ in \twodba\ can be removed.

\subsec{ZZ-FZZT and ZZ-ZZ annulus amplitudes}

Now it is almost trivial to obtain from \annulusfin\ the ZZ-FZZT
annulus amplitude, by using the fact that the ZZ branes can be
written as a difference of two FZZT branes (this fact follows from
\refs{\ZamolodchikovAH, \HosomichiXC, \PonsotSS} and was made most
explicit in \MartinecKA). To obtain the correct ZZ-FZZT annulus
amplitude, however, we have to take into account one subtlety,
concerning the normalization of the ZZ boundary state. This is
discussed in appendix B, and the result is that the ZZ branes of
minimal string theory come with an extra minus sign relative to
the ZZ branes of Liouville theory alone:
 \eqn\ZZdiff{|m,n\rangle = |\sigma=\sigma_{m,-n}\rangle -
 |\sigma=\sigma_{m,n}\rangle } with \eqn\sigmadef{ \sigma_{m,n} =
 i\left({m\over b}+n b\right)\ . }
Therefore the annulus amplitude between an $(m,n)$ ZZ brane and an
FZZT brane labelled by $\sigma$ is simply
 \eqn\annulusZZFZZT{Z(m,n|\sigma) = Z(\sigma_{m,-n}|\sigma)
 -Z(\sigma_{m,n}|\sigma) =\log\left({z-z_{m,n}^-\over z-z_{m,n}^+}
 \right)}
where $z_{m,n}^{\pm}$ was defined in \introzmn. In deriving
\annulusZZFZZT, we have used the fact that
$T_p(z_{m,n}^{+})=T_{p}(z_{m,n}^{-})$. The fact that the $x$
dependence cancels out in the ZZ-FZZT annulus amplitude agrees
with the idea that this dependence comes from the contribution of
zero-area surfaces to the path integral. Since worldsheets with ZZ
branes have boundaries of infinite length, they cannot give rise
to zero-area surfaces.

As in \loopinsert-\twodba\ we can exponentiate \annulusZZFZZT\ to
find the expectation value of the FZZT brane in the presence of a
ZZ brane.  In the one matrix model
 \eqn\annulusZZFZZTe{\langle \det(x-M)\rangle_{ZZ_n}
  ={z-z_{1,n}^-\over z-z_{1,n}^+} \ (1+\CO(1/N)).}
This expression vanishes at $z=z^-_{1,n}$ which is the position of
the ZZ brane in the first sheet.  (It diverges at $z=z^+_{1,n}$
which is in the second sheet.)  This is consistent with the
interpretation of the ZZ brane as an eigenvalue in the first sheet
at $z^-_{1,n}$, which makes the expectation value of $\det(x-M)$
vanish.

Although we have specialized (without loss of generality) to the
matter state $(1,1)$ in our discussion so far, it is instructive
to consider the ZZ-FZZT annulus amplitudes for other combinations
of Liouville and matter labels. One way to do it is by using the
expressions in appendix A.  Interestingly, the following four
choices of ZZ boundary states lead to the same annulus amplitude
\annulusZZFZZT: \eqn\LmnMmn{
Z(m,n;1,1|\sigma;1,1)=Z(1,1;m,n|\sigma;1,1)
=Z(1,n;m,1|\sigma;1,1)= Z(m,1;1,n|\sigma;1,1)\ . } This degeneracy
of ZZ branes was also observed in the instanton effects of the
two-matrix model \KazakovDU. It can be derived using the boundary
state identifications (3.7)--(3.8) of \SeibergNM.

To complete the discussion, let us consider the ZZ-ZZ annulus amplitude.
Using the relation \ZZdiff\ once again, it can be obtained from the ZZ-FZZT
annulus amplitude \annulusZZFZZT:
 \eqn\annulusZZZZ{Z(m,n|m^\prime,n^\prime) =
 Z(m,n|\sigma_{m^\prime,-n^\prime})-
 Z(m,n|\sigma_{m^\prime,n^\prime}) = \log {(z^+_{m,n} -
 z^+_{m^\prime,n^\prime})(z^-_{m,n} -
 z^-_{m^\prime,n^\prime}) \over (z^+_{m,n} -
 z^-_{m^\prime,n^\prime})(z^-_{m,n} - z^+_{m^\prime,n^\prime}) } \ . }
Note that the annulus amplitude diverges for $(m,n)=(m',n')$.  To
get more insight into this divergence we exponentiate
\annulusZZZZ\ as above to find the partition function of two ZZ
branes
 \eqn\twoZZ{{(z^+_{m,n} - z^+_{m^\prime,n^\prime})(z^-_{m,n} -
 z^-_{m^\prime,n^\prime}) \over (z^+_{m,n} -
 z^-_{m^\prime,n^\prime})(z^-_{m,n} - z^+_{m^\prime,n^\prime})}\ .}
Note that unlike \annulusZZFZZTe, which has a single zero at
$z=z^-_{m,n}$, here we have a double zero for $(m,n)=(m',n')$.
This is in agreement with the interpretation of the ZZ branes as
eigenvalue.  The double zero in \twoZZ\ is the standard double
zero for two equal eigenvalues in the matrix model.

We will return to this divergence of \annulusZZZZ\ and the
corresponding zero of \twoZZ\ below.

\subsec{Target space picture}

In section 2.1 we considered the FZZT-FZZT annulus amplitude in
$\ell$-space \refs{\MooreIR,\MooreAG}, which is an important
source of physical intuition.  Here we consider the analogous
transform of the ZZ-FZZT annulus amplitude \annulusZZFZZT.  This
``minisuperspace amplitude" of the ZZ brane is defined to be the
function $\Psi_{m,n}(\ell)$ that satisfies \eqn\ZZwvfndef{
Z(m,n|\sigma) = -\int_{0}^\infty{d\ell\over \ell}e^{-\ell\cosh\pi b
\sigma}\Psi_{m,n}(\ell)\ . } This identifies the loop length
$\ell$ with the integral of the boundary cosmological constant
operator \MartinecHT, \eqn\ellident{ \ell \leftrightarrow \oint
e^{b\phi}\ . } In the minisuperspace approximation, this is just
$e^{b\phi_0}$ for some constant $\phi_0$, and therefore we can
think of $\Psi_{m,n}(\ell=e^{b\phi_0})$ as measuring the
fluctuations in the target space fields induced by placing a ZZ
brane at $\phi\to+\infty$.

Note that we have avoided calling $\Psi_{m,n}(\ell)$ a
wavefunction for the ZZ brane, although its definition is similar
to that of the wavefunctions of local closed-string operators. The
reason it is inaccurate to think of $\Psi_{m,n}(\ell)$ as a
wavefunction is because the ZZ brane is not a local puncture on
the worldsheet, but rather a macroscopic boundary. In fact, as we
noted above, the boundary of a worldsheet ending on a ZZ brane has
infinite length.

Now let us derive a compact formula for $\Psi_{m,n}(\ell)$ and
discuss some of its properties. Although in general
the inverse Laplace transform can be a difficult operation, here
our task is greatly aided by the fact that the annulus amplitude
\annulusZZFZZT\ satisfies a rather simple differential equation:
\eqn\anndiffeq{
(x-x_{mn})\partial_x Z(m,n|\sigma) = {4\over
p}\sum_{k=1}^{p-1}{\sinh{ \pi k \sigma \over \sqrt{pq}}\over
\sinh{\pi p\sigma\over\sqrt{pq}}}\ \sin{\pi m(p-k)\over p}\
\sin{\pi n(p-k)\over q} ~.
}
Since the Laplace transform of the Bessel function $K_{k\over
p}(\ell)$ is
\eqn\LaplaceK{
\int_0^\infty d\ell e^{-\ell\cosh \pi b
\sigma}{1\over\pi}K_{k\over p}(\ell)\sin{\pi k\over p} ={\sinh{
\pi k \sigma \over \sqrt{pq}}\over \sinh{\pi
p\sigma\over\sqrt{pq}}}~,
}
we see that \anndiffeq\ becomes a first-order differential
equation for $\Psi_{m,n}(\ell)$:
\eqn\derivpsi{
{\partial\over\partial\ell}\Psi_{m,n}(\ell) =
x_{mn}\Psi_{m,n}(\ell) + J_{m,n}(\ell)
}
where
\eqn\Jmnz{
J_{m,n}(\ell)={4\over p}\sum_{k=1}^{p-1} K_{k\over
p}(\ell)\sin{\pi k\over p}\sin{\pi m(p-k)\over p}\sin{\pi
n(p-k)\over q}~.
}
This equation can be easily solved, once we impose the boundary
condition $\Psi_{m,n}(0)=0$, which follows from the fact that
$\lim_{x\to\infty}\partial_x Z(m,n|\sigma)=0$. Then we find
\eqn\Psiellint{
\Psi_{m,n}(\ell)=\int_0^{\ell} d\ell'\,
\Psi_{m,n}^{(0)}(\ell-\ell') J_{m,n}(\ell')
}
where
\eqn\Psizero{
\Psi_{m,n}^{(0)}(\ell)=e^{x_{m,n}\ell}
}
is a solution to \derivpsi\ with $J_{m,n}=0$.

We interpret \Psiellint\ as follows. If the minisuperspace
amplitude $\Psi_{m,n}$ were equal to $\Psi^{(0)}_{m,n}$ for all
$\ell$, the annulus amplitude would be given by
\eqn\naiveann{
Z(m,n|\sigma) \sim -\int_{\Lambda^{-1}}^\infty
{d\ell\over\ell}e^{-x\ell}
\Psi^{(0)}_{m,n}(\ell)\sim \log\left({x-x_{m,n}\over\Lambda}\right)
}
where $\Lambda^{-1}$ is a short $\ell$ cutoff. (Note that this
short $\ell$ cutoff is not necessary in the Laplace transform of
the correct minisuperspace amplitude $\Psi_{m,n}(\ell)$, since
this function vanishes at $\ell=0$.) Given that the FZZT brane
corresponds to $\Tr\log(x-X)$ in the dual matrix model, \naiveann\
amounts to naively replacing the matrix $X$ with the eigenvalue
$x_{m,n}$. Then the modification to \naiveann\ due to the source
term $J_{m,n}$ can be interpreted as the effect of interactions
between the ZZ brane and other eigenvalues in the Fermi sea.

Since $\Psi_{m,n}(\ell=e^{b\phi_0})$ probes the target-space
location of the ZZ brane, we can use the large and small $\ell$
behavior of $\Psi_{m,n}(\ell)$ to test the idea that the ZZ branes
are located at $\phi\to+\infty$ in target space. Let us consider
the two limits separately:

\lfm{1.} At small $\ell$, we use the fact that $K_{k\over
p}(\ell)\sim \ell^{-{k\over p}}$ to find
\eqn\smallpsi{
\Psi_{m,n}(\ell)\to {2^{1-{1\over p}}(z_{m,n}^{-}-z_{m,n}^{+})
\over\Gamma\Big({1\over p}\Big)} \,\ell^{1\over p}\qquad
(\ell\to0) ~.
}
The vanishing of $\Psi_{m,n}(\ell)$ at $\ell=0$ implies that the
ZZ branes are {\it not} located at $\phi\to-\infty$ in the weak
coupling region of the Liouville direction. Contrast this with the
small $\ell$ behavior of the wavefunctions of local closed-string
operators \MooreIR\ -- these always behave as $\ell^{-|\nu|}$ as
$\ell\to0$, since local operators are concentrated at
weak-coupling \SeibergEB.

\lfm{2.} From the explicit form of $J_{m,n}(\ell)$ \Jmnz, it is
not so difficult to show that
\eqn\Jmnint{
\int_0^\infty d\ell \,e^{-x_{m,n}\ell}J_{m,n}(\ell)=1 ~,
}
which implies that at large $\ell$,
\eqn\largelpsi{
\Psi_{m,n}(\ell)\to \exp\big(x_{m,n}\ell\big) \qquad
(\ell\to\infty) ~.
}
Before we interpret \largelpsi, we must take into account the fact
that the FZZT brane dissolves at strong coupling and therefore
cannot necessarily probe the ZZ brane located there. To remove
the effect of the dissolving FZZT brane, it is reasonable to
divide the minisuperspace amplitude by, say, the FZZT disk
amplitude $Z(\ell)$. Since this vanishes as $K_{\nu}(\ell) \sim
e^{-\ell}$ as $\ell\to\infty$, the ratio of $\Psi_{m,n}(\ell)$ to
$Z(\ell)$ is always exponentially increasing at large $\ell$
(recall that $|x_{mn}|<1$). This confirms that the ZZ branes are
indeed localized deep in the strong coupling region.

\newsec{Geometric Interpretation}

\subsec{ZZ brane deformations}

Having worked out in detail the properties of the ZZ-FZZT annulus
amplitude in the previous section, let us now interpret this
quantity geometrically. In \SeibergNM, it was argued that adding
$(m,n)$ ZZ branes to the background splits the $(m,n)$
singularity, giving a nonzero value to the integral of $y\,dx$
around the singularity. Now we are in a position to understand
this explicitly using the ZZ-FZZT annulus amplitude. The important
thing to note is that, to leading order, the change in the FZZT
disk amplitude due to the addition of $N_{mn}$ background ZZ
branes is measured by the annulus amplitude:
\eqn\FZZTZZdef{
\delta_{m,n} Z(\sigma) = g_s N_{m,n} Z(m,n|\sigma) ~.
}
Here we must require $g_s N_{mn}\ll 1$ in order for the
perturbation expansion to make sense. A deformation of the disk
amplitude leads to a deformation of the curve $y=\partial_x
Z(\sigma)$,
\eqn\FZZTZZdefII{
\delta_{m,n} y = \partial_x\delta_{m,n} Z \sim
\partial_x Z(m,n|\sigma) = {1\over p U_{p-1}(z)}\left({1\over z-z_{m,n}^-}-{1\over
z-z_{m,n}^+}\right)~,
}
and finally to a deformation of the surface
\eqn\FZZTZZdefsub{
\delta_{m,n} F = p U_{p-1}(y)\delta y \sim {U_{p-1}(T_q(z))\over
U_{p-1}(z)}\left({1\over z-z_{m,n}^-}-{1\over
z-z_{m,n}^+}\right)~.
}
It is not too difficult to see that $\delta_{m,n} F$ is a
polynomial in $z$. It turns out that it is also a polynomial in
$x$ and $y$:
\eqn\FZZTZZdefxy{
\delta_{m,n}F\sim -{8(-1)^{m+n}\sin{\pi np\over q}\sin{\pi mq\over
p} }\sum_{r,s} U_{s-1}(x_{mn})U_{r-1}(y_{mn})
U_{s-1}(x)U_{p-r-1}(y)
}
where the sum runs over the range $1\le r\le p-1$, $1\le s\le
q-1$, $qr-ps>0$. The easiest way to see that \FZZTZZdefsub\ and
\FZZTZZdefxy\ are equal is to notice that both evaluate to the
same values at the $(p-1)(q-1)/2$ singularities of $\CM_{p,q}$:
\eqn\FZZTZZdefsing{
\delta_{m,n} F(x_{m'n'},y_{m'n'}) \sim \left({(-1)^{(q+1)m+(p+1)n}
pq\over \sin{\pi np\over q}\sin{\pi mq\over
p}}\right)\delta_{m,m'}\delta_{n,n'} ~.
}
This means that, in terms of $z$, they actually coincide at the
$(p-1)(q-1)$ different points $z_{m'n'}^{\pm}$. Since they are
both degree $pq-p-q-1$ polynomials in $z$, the fact that they
coincide at $(p-1)(q-1)$ different points implies that they must
in fact be equal everywhere.

In the process of showing that \FZZTZZdefsub\ and \FZZTZZdefxy\
are equal, we have also shown which singularities of $\CM_{p,q}$
are split by $\delta_{m,n}F$. According to \FZZTZZdefsing, the
$(m,n)$ ZZ brane deformation vanishes at every singularity except
the $(m,n)$ singularity. But we also know that a deformation
preserves a given singularity if and only if it vanishes on that
singularity. Therefore, the effect of adding $(m,n)$ ZZ branes to
the background is to split precisely the $(m,n)$ singularity,
leaving the others unchanged. In other words, the ZZ branes
``diagonalize" the $A$-cycle deformations of $\CM_{p,q}$. This
confirms very explicitly the arguments in \SeibergNM.

Finally, we should discuss the effect of the ZZ branes on the
$B$-cycles of $\CM_{p,q}$. Recall that the integral of $y\,dx$
around the $(m',n')$ $B$-cycle of the surface corresponds to the
$(m',n')$ ZZ brane action,
 \eqn\ydxmnB{ \oint_{B_{m',n'}}y\,dx =
 Z_{m',n'} ~.
 }
Adding ZZ branes deforms this period in two ways: first, it
deforms the curve $y$ by $\delta_{m,n}y$ given in \FZZTZZdefII;
and second, it deforms the contour of integration $B_{m',n'}$. For
$(m',n')\ne (m,n)$, the latter effect is subleading, in which case
the leading-order deformation to the period is just the annulus
amplitude \annulusZZZZ:
 \eqn\ydxmnBdefZZ{
 \delta_{m,n}\oint_{B_{m',n'}}y\,dx = g_s N_{m,n} Z(m,n|m',n'),
 \qquad (m',n')\ne (m,n) ~.
 }
However, for $(m',n')=(m,n)$ the deformation to the contour of
integration $B_{m,n}$ is important, because of the pole in
$\delta_{m,n}y$ at $x_{mn}$. Since the ZZ deformation also splits
the singularity at $x_{mn}$ into a branch cut whose width is a
positive power of $g_s N_{m,n}$, this has the effect of cutting
off the integral $\delta_{m,n} \oint_{B_{m,n}}y\,dx$ at an
infinitesimal distance from the pole. Therefore, the diagonal
deformation is given by
\eqn\ydxmnBdefZZdiag{
\delta_{m,n} \oint_{B_{m,n}}y\,dx \sim g_s N_{m,n}\log (g_s
N_{m,n}) ~.
}
Note that if we had not cut off the integral near the singularity,
we would have found a logarithmic divergence. This corresponds to
the divergence in the annulus amplitude \annulusZZZZ\ for
$(m',n')=(m,n)$. We should also point out that the logarithmic
enhancement for $(m',n')=(m,n)$ agrees with the matrix model
interpretation of the ZZ branes as fermionic eigenvalues, as
discussed at the end of section 2.2. If we try to insert multiple
ZZ branes at the same point $x_{mn}$, they will repel one another
due to their Fermi statistics, leading to an enhanced backreaction
on the surface at $x_{mn}$.

Since \ydxmnBdefZZ\ and \ydxmnBdefZZdiag\ are always non-zero, we
conclude that the ZZ branes deform all the $B$-cycles of
$\CM_{p,q}$. However, there is a sense in which the ZZ branes
``diagonalize" the deformations of the $B$-cycles, since the
$(m,n)$ deformation to the $(m,n)$ $B$-cycle receives a
logarithmic enhancement relative to the other cycles.

\subsec{Tachyon eigenstate deformations}

We have just seen that adding $(m,n)$ ZZ branes deforms the
$(m,n)$ $A$-cycle of $\CM_{p,q}$, leaving the other $A$-cycles
unchanged. In this subsection, we will answer a closely related
question: what are the deformations that deform just the $(m,n)$
$B$-cycle of $\CM_{p,q}$, leaving the other cycles unchanged?
Since these deformations preserve the $A$-cycles, they must be
linear combinations of the singularity-preserving deformations,
i.e.\ they correspond to physical closed-string operators. Then
following the reasoning that led to \ydxmnBdefZZ, we find that the
leading-order effect of adding a closed-string operator
$\CV_{r,s}$ to the worldsheet action is to deform the $(m,n)$
$B$-cycle by the $(m,n)$ ZZ one-point function of the operator:
\eqn\ydxmnBdef{
\delta_{r,s} \oint_{B_{m,n}}y\,dx \sim \langle
\CV_{r,s}|m,n\rangle \sim \sin {\pi r (m q+np)\over p}\ \sin {\pi
s (mq+n p)\over q} ~.
}
By forming linear combinations of the $\CV_{r,s}$, we can clearly
arrange for only one particular $B$-cycle to be deformed.

An especially interesting example is when the $\CV_{r,s}$ are the
physical tachyon operators $\CT_{r,s}$ at ghost number one. To
find the linear combinations of tachyons that diagonalize the
deformations, recall that the tachyons form a module under the
action of the ground ring. Thus we can find linear combinations
$\tilde\CT_{m,n}$ of the tachyons that are eigenstates of the
ring,
\eqn\tacheigenII{\eqalign{
 \hat x \tilde\CT_{m,n} &= x_{mn}\tilde\CT_{m,n} ~,\cr
 \hat y \tilde\CT_{m,n} &= y_{mn}\tilde\CT_{m,n} ~.\cr
}}
Here $\hat x=\hat\CO_{1,2}/2$ and $\hat y=\hat\CO_{2,1}/2$ denote
the generators of the ground ring. Since the ZZ branes are also
eigenstates of the ring, we find
\eqn\tacheigenIII{
\langle \hat x \tilde\CT_{m,n}|m',n'\rangle = x_{mn} \langle
\tilde\CT_{m,n}|m',n'\rangle = x_{m'n'} \langle
\tilde\CT_{m,n}|m',n'\rangle
}
and similarly for $\hat y$. This is, of course, only possible if
and only if
\eqn\tacheigenfin{
\langle\tilde\CT_{m,n}|m',n'\rangle\propto
\delta_{m,m'}\delta_{n,n'}\ .
}
In other words, the tachyon eigenstates diagonalize the
deformations of the $B$-cycles of $\CM_{p,q}$, with the $(m,n)$
eigenstate deforming only the $(m,n)$ $B$-cycle.

Note that in this argument, we have implicitly used the fact that
the tachyon eigenstates and the ZZ branes have the same set of
eigenvalues. This follows from the ring relations $U_{q-1}(\hat
x)=U_{p-1}(\hat y)=0$, together with the additional relation in
the tachyon module $T_{p}(\hat y)-T_q(\hat x)=0$. Since this
latter relation is also the equation for $\CM_{p,q}$, the
eigenvalues in the tachyon module must coincide with the
singularities of $\CM_{p,q}$, which are of course the eigenvalues
of the ZZ branes.

Finally, let us find an explicit formula for the tachyon
eigenstates in terms of the usual basis $\CT_{r,s}$. This becomes
trivial, once one realizes that because the modular transformation
matrix $S$ diagonalizes the fusion rules, it also diagonalizes the
ring elements and their action on the tachyon module. Since the
modular $S$ matrix can be written
\eqn\Smat{
S_{(m,n),(r,s)} = S_{(m,n),(1,1)} U_{s-1}(x_{mn})U_{r-1}(y_{mn})
=-\sqrt{8\over pq}(-1)^{sm+rn}\sin {\pi snp\over q}\sin{\pi
rmq\over p}~,
}
the tachyon eigenstates are given by
\eqn\tacheigen{
\tilde\CT_{m,n} = \sum_{r,s}
U_{s-1}(x_{mn})U_{r-1}(y_{mn})\CT_{r,s} ~.
}
Of course, one can also prove this formula in a more
straightforward way, by acting on \tacheigen\ with $\hat
x-x_{mn}$. For this, one needs to use the ring relations, together
with the fact that the tachyons can be written in terms of the
ground ring, $\CT_{r,s}=U_{s-1}(\hat x)U_{r-1}(\hat y)\CT_{1,1}$.
The recurrence relation $xU_{r-1}(x) = U_{r}(x)+U_{r-2}(x)$ also
comes in handy.

To summarize, we have shown that the $(m,n)$ tachyon eigenstate
deforms only the $(m,n)$ $B$-cycle of $\CM_{p,q}$, leaving the
other $B$-cycles unchanged. Comparing with our results on the ZZ
brane deformations, we see hints of an interesting duality
emerging between the tachyon eigenstates and the ZZ branes -- they
are both eigenstates of the ring, and while the former
diagonalizes the $B$-cycles, the latter diagonalizes the
$A$-cycles of $\CM_{p,q}$. We will discuss in detail the evidence
for this duality in section 4. In the meantime, let us first wrap
up a loose end in the analysis of the ZZ brane deformations.

\subsec{General discussion of the deformations}

In section 3.1, we found the deformations to the surface due to
adding background ZZ branes. But there is one point that we
neglected to address. One might have wondered whether the ZZ brane
deformations span the entire space of polynomial,
singularity-destroying deformations, or whether there are other
such deformations that do not correspond to ZZ branes. To answer
this question, we need to study the polynomial deformations of
$\CM_{p,q}$ in slightly more generality.

Consider the space of polynomial, singularity-destroying
deformations of $\CM_{p,q}$. By this, we mean the quotient ring
\eqn\cosetdef{
    \CP^-\equiv {\Bbb C[x,y]\over\{F\}\cup \CP^+}
}
of polynomials in $x$ and $y$ modulo the original curve
$F=T_p(y)-T_q(x)$ and modulo the ideal
\eqn\Ppdef{
\CP^+ = \{\ U_{p-r-1}(y)U_{s-1}(x)-U_{r-1}(y)U_{q-s-1}(x)\ \}
}
of polynomial, singularity-preserving deformations of the curve.
We claim that the deformations
\eqn\defsimp{
\tilde\delta_{r,s} F = U_{s-1}(x)U_{p-r-1}(y), \qquad 1\le r\le
p-1,\quad 1\le s\le q-1,\quad qr-ps>0
}
form a complete basis for $\CP^-$. A sketch of the proof is as
follows. First, consider the most general polynomial in $x$ and
$y$. Using $F=0$, we can clearly reduce any such polynomial down
to one whose degree in $y$ is $\le p-1$. Since the Chebyshev
polynomials are a complete set of orthogonal polynomials, this
means that
\eqn\basisgenII{\eqalign{
U_{k-1}(x)U_{j-1}(y),\qquad 1\le j\le p,\qquad k\ge 1
}}
is a basis for $\Bbb C[x,y]/\{F\}$. Now, consider the effect of
modding out by $\CP^+$. From the form \Ppdef\ of the
singularity-preserving deformations, we see that this essentially
means that we have the freedom to reflect $(j,k)\to (p-j,q-k)$. If
$j=p$ or $k=q$, this is clearly an element of $\CP^+$. If $1\le
k\le q-1$, either \basisgenII\ or its reflection is of the form
\defsimp. Finally, if $k>q$, a reflection turns \basisgenII\ into
$U_{q-k-1}(x)U_{p-j-1}(y) = -U_{k-q-1}(x)U_{p-j-1}(y)$. Applying
this transformation repeatedly, we can reduce $k$ down to the
range $1\le k\le q-1$, where the previous case holds. This proves
that \defsimp\ form a complete basis for the
singularity-destroying deformations of $\CM_{p,q}$.

According to \defsimp, the number of singularity-destroying
deformations is the same as the number of singularities of
$\CM_{p,q}$. Since there are also this many ZZ brane deformations,
these must form an equivalent basis for the space of polynomial,
singularity-destroying deformations of $\CM_{p,q}$. Indeed, it is
immediately obvious from \FZZTZZdefxy\ that the ZZ brane
deformations are related to the basis \defsimp\ by the action of
the modular $S$-matrix \Smat.

\newsec{ZZ Branes as ``Wrong-branch" Tachyons}

We have seen that the ZZ branes and the tachyon eigenstates have
very analogous effects on the surface. The former deforms the
$(m,n)$ $A$-cycle of $\CM_{p,q}$, while the latter deforms the
$(m,n)$ $B$-cycle. This suggests that the two are in some sense
conjugate or dual to one another. In this section, we will try and
make this duality more precise.

We propose that the ZZ branes can be thought of as ``wrong-branch"
tachyons. Before we proceed to discuss the evidence for our
proposal, perhaps it will be useful to recall the definition of
wrong-branch tachyons. These are tachyons whose Liouville part
$e^{2\alpha\phi}$ violates the bound $\alpha<Q/2$ \SeibergEB. They
are usually excluded on the basis that they do not correspond to
local closed-string operators. One way to see this is by studying
the semiclassical wavefunctions in the minisuperspace
approximation. Tachyons with $\alpha<Q/2$ have wavefunctions
$K_{\nu}(\ell)$ with
\eqn\nudef{
\nu = {2\alpha-Q\over b}
}
that are localized at small $\ell$ and decay exponentially at
$\ell\to\infty$ \MooreIR. This is the expected behavior for a
local operator inserted at weak coupling. On the other hand,
tachyons with $\alpha>Q/2$ have the opposite behavior.
Their wavefunctions are given by $I_{|\nu|}(\ell)$, which vanish
at $\ell=0$ and grow exponentially at large $\ell$.

Another way to see that the wrong-branch tachyons do not
correspond to local closed-string operators is in the target-space
picture, where $\ell\sim e^{b\phi}$. Writing the right-branch
wavefunction as $K_{\nu}(\ell)\sim
I_{-|\nu|}(\ell)-I_{|\nu|}(\ell)$, we see that the wavefunction
can be thought of as the superposition of an incoming mode
$I_{-|\nu|}$ propagating towards $\phi\to+\infty$, and an outgoing
mode $I_{|\nu|}$ describing reflection off the Liouville potential
$e^{2b\phi}$. Then since the wrong-branch wavefunctions are just
$I_{|\nu|}(\ell)$, they describe outgoing modes, with no
corresponding incoming mode. Thus they do not describe insertions
of local operators in the weak-coupling region. Imagine, however,
the effect of placing a ZZ brane at $\phi\to+\infty$. This would
presumably emit closed-string radiation in the $\phi\to-\infty$
direction, i.e.\ it would give rise to outgoing modes with no
incoming counterpart. This is our first hint that the ZZ branes
are related to the wrong-branch tachyons.

Now let us discuss in detail the evidence for a correspondence
between the ZZ branes and the wrong-branch tachyons. First, notice
the similarity between the form of the ZZ brane deformations
\FZZTZZdefxy\ and the tachyons \tacheigen, both of which are
eigenstates of the ground ring. This suggests that there is a
sense in which the $(m,n)$ ZZ brane is a linear combination of
wrong-branch $(r,s)$ tachyons with well-defined KPZ scaling (we
will discuss the KPZ scaling below), i.e.\
\eqn\wrongbranch{
|m,n\rangle = \sum_{r,s\atop
qr-ps>0}U_{s-1}(x_{mn})U_{r-1}(y_{mn})|r,s\rangle\rangle
}
by analogy with \tacheigen. (We should think of \wrongbranch\ as
defining the states $|r,s\rangle\rangle$, i.e.\ these states are
special linear combinations of the usual $(m,n)$ ZZ branes.) We
can also provide a simpler explanation for \wrongbranch. Recall
that the $(m,n)$ ZZ brane with  matter state $(1,1)$ can also be
thought of as a $(1,1)$ ZZ brane with matter state $(m,n)$. The
linear transformation between the $(m,n)$ matter Cardy states and
the $(r,s)$ Ishibashi states is essentially the matrix that
appears in \wrongbranch.\foot{More precisely, the matter Cardy and
Ishibashi states are related via the matrix
${S_{(m,n),(r,s)}\over\sqrt{S_{(1,1),(r,s)}}}$, with $S$ given in
\Smat. The difference between this and \wrongbranch\ can be
absorbed into the normalizations of $|r,s\rangle\rangle$ and
$|m,n\rangle$.} Thus we can think of the states
$|r,s\rangle\rangle$ as labelling the different matter Ishibashi
states. In this sense, they have well-defined closed-string
quantum numbers (e.g.\ KPZ dimension).

The second piece of evidence for our proposal comes from the ZZ
brane minisuperspace amplitudes discussed in section 2.3. There we
showed that they vanish at $\ell=0$ as $\ell^{1/p}$ and grow
(relative to the FZZT disk amplitude) exponentially at large
$\ell$. This agrees with the expected behavior of wrong-branch
wavefunctions. We can actually make a much more detailed test of
our proposal by expanding the minisuperspace amplitude to higher
orders in $\ell$ at small $\ell$. This corresponds to the
semiclassical regime (i.e.\ weak coupling), where we expect to
find the Bessel functions $I_{|\nu|}(\ell)$. Indeed, at large $x$
the annulus amplitude has the expansion
\eqn\esisin{
Z(m,n|\sigma)=-\sum_{k=1}^\infty{4\over
k}e^{-k{\pi\sigma\over\sqrt{pq}}} \sin{\pi mk\over p}\sin{\pi
nk\over q} ~.
}
The Laplace transform of this expression can be found by using the
formula
\eqn\LapI{
\int_0^\infty{d\ell\over\ell}e^{-\ell\cosh\tau}I_{\nu}(\ell)
={1\over \nu}e^{-\nu\tau}.
}
{}From \esisin\ and \LapI, we obtain the minisuperspace amplitude
$\Psi_{m,n}(\ell)$ at small $\ell$ as an infinite sum of modified
Bessel functions
\eqn\Psimn{
\Psi_{m,n}(\ell)={4\over p}\sum_{k=1}^\infty I_{k\over p}(\ell)
\sin{\pi mk\over p}\sin{\pi nk\over q} ~.
}
The physical interpretation of this expansion is clear: these are
the fluctuations in the closed-string modes at $\phi\to-\infty$
due to the insertion of an $(m,n)$ ZZ brane at $\phi\to+\infty$.
The fluctuations are all described by the Bessel functions
$I_{\nu}$, and therefore they are all on the wrong-branch.\foot{It
is also nice that the terms with $k=0\ \mod\ p, q$ do not
contribute to the expansion. This is consistent with the BRST
analysis of Lian and Zuckerman \LianGK.}

The third and final piece of evidence for our proposal was alluded
to above, namely the KPZ scaling of the ``ZZ branes"
$|r,s\rangle\rangle$. Again, we can extract this from the small
$\ell$ behavior of the minisuperspace amplitude; if
$\Psi_{m,n}(\ell)$ behaves as $\ell^{\nu}$ as $\ell\to 0$, then
the KPZ dimension $\alpha$ of the associated operator is given in
\nudef. For instance, the tachyons have $\nu=-(qr-ps)/p$, so
$\alpha = {p+q-qr+ps\over 2\sqrt{p q}}$, which is the correct
formula for the Liouville exponent of the tachyon. According to
\wrongbranch\ and \FZZTZZdefxy, the annulus amplitude of the state
$|r,s\rangle\rangle$ with an FZZT brane is
\eqn\opfrsZZ{
\partial_x Z(r,s|\sigma) = {U_{p-r-1}(y)U_{s-1}(x)\over
U_{p-1}(y)} ~.
}
Expanding this at large $\sigma$, we find to leading order that
\eqn\opfrsZZexpand{
Z(r,s|\sigma)\sim e^{-\left({qr-ps\over p}\right)\pi b \sigma} ~.
}
Therefore, the minisuperspace amplitude behaves as
\eqn\psirs{
    \Psi_{r,s}(\ell) \sim I_{qr-ps\over p}(\ell) \sim
    \ell^{qr-ps\over p}
}
at small $\ell$. Then according to \nudef, the KPZ dimension of
the operators $\CZ_{r,s}$ is
\eqn\KPZrs{
\alpha_{r,s} = {p+q+qr-ps\over 2\sqrt{p q}}
}
which is precisely the expected dimension for a wrong-branch
$(r,s)$ tachyon.

\newsec{A Closer Look at the $(p,q)=(2,2k-1)$ Theories}

Here we will discuss in detail the models with $(p,q)=(2,2k-1)$.
Apart from serving as concrete examples for the general analysis
above, these models also contain many additional simplifications.
For instance, it is easy to see that the minisuperspace amplitude
studied above becomes particularly simple when $p=2$. The source
term $J_{1,n}$ in \Jmnz\ is not a sum of many Bessel functions,
but is given by a single function $K_{1\over 2}(z)=\sqrt{\pi\over
2z}e^{-z}$. Also, we can set $m=1$ for the $(m,n)$ ZZ-brane, since
the independent set of ZZ-branes is spanned by $(m,n)=(1,n)$ with
$n=1,\cdots,(q-1)/2$ \SeibergNM. Then the minisuperspace amplitude
of the $(1,n)$ ZZ brane \Psiellint\ is just
\eqn\onematpsi{
\Psi_{1,n}(\ell)=\exp\Big(x_{1,n}\,\ell\Big) {\rm
Erf}\left(\sin{\pi n\over q}\,(2\ell)^{1\over2}\right)
}
where Erf$(z)$ is the error function defined by
\eqn\Efrdef{
{\rm Erf}(z)={2\over\sqrt{\pi}}\int_0^zdx\,e^{-x^2}
}
and $x_{1,n}$ is given by
\eqn\xonenn{
x_{1,n}=-\cos{2\pi n\over q} ~.
}

One of the reasons for the simplification at $p=2$ is that the
$(2,q)$ theory is described by a one-matrix model. Accordingly,
the Riemann surface ${\cal M}_{2,q}$ for this theory is a double
cover of the $x$-plane \SeibergNM
\eqn\ptwocurve{
F(x,y)=T_q(x)+1-2y^2=0 ~.
}
On the $x$-plane, there is a square root cut along $x\leq -1$ and
the points $x=x_{1,n}$ are the singularities of the curve
\ptwocurve.

The minisuperspace amplitude \onematpsi\ can be interpreted more
geometrically if we write it as a line integral on this curve. For
this purpose, it is useful to rewrite the amplitude
\annulusZZFZZT\ in terms of the $x$-coordinate
\eqn\Annptwo{
Z(1,n|\sigma)=\log\left({\sqrt{x+1}-\sqrt{x_{1,n}+1}\over\sqrt{x+1}+\sqrt{x_{1,n}+1}}\right)
}
where we substituted the following relations valid for the $p=2$
case
\eqn\zonenpm{
z=\sqrt{x+1\over2},\quad z_{1,n}^{\pm}=\mp\sin{\pi n\over
q}=\mp\sqrt{x_{1,n}+1\over 2} ~.
}
Then the minisuperspace amplitude in question is just an inverse
Laplace transform of
\eqn\delxann{
\partial_xZ(1,n|\sigma)={\sqrt{x_{1,n}+1}\over\sqrt{x+1}(x-x_{1,n})}.
}
The usual prescription for the inverse Laplace transform of a
function $f(x)$ is to integrate $e^{\ell x}f(x)$ along a contour
parallel to the imaginary $x$-axis and to the right of all the
singularities of $f(x)$. Applying this to \delxann\ results in the
following formula for $\Psi_{1,n}(\ell)$:
\eqn\psiAcycle{
\Psi_{1,n}(\ell)=\oint_{C} {dx\over2\pi i}\, e^{\ell x}
{\sqrt{x_{1,n}+1}\over\sqrt{x+1}(x-x_{1,n})}
}
where $C$ denotes a contour surrounding the cut $x\leq-1$ and the
pole $x=x_{1,n}$.

As discussed in \HanadaIM, we can gain more insight into the
effect of ZZ-brane by introducing the density $\rho_{1,n}(x)$ as
\eqn\rhomndef{
\Psi_{1,n}(\ell)=\int_{-\infty}^\infty dx\,e^{\ell
x}\rho_{1,n}(x) ~.
}
$\rho_{1,n}(x)$ can be thought of as the deformation of the
eigenvalue density due to the extra eigenvalue introduced at
$x=x_{1,n}$. According to \psiAcycle, it is given by the imaginary
part of $-\partial_xZ(1,n|\sigma)$ across the real $x$-axis. The
explicit form of $\rho_{1,n}(x)$ is found to be
\eqn\rhoform{
\rho_{1,n}(x)=\delta(x-x_{1,n})-
\theta(-1-x){1\over\pi}{\sqrt{1+x_{1,n}}\over\sqrt{-1-x}(x_{1,n}-x)} ~.
}
Here $\theta(-1-x)$ denotes the step function, which is 1 for
$x\leq -1$ and 0 for $x>-1$. The $\delta$-function in \rhoform\
represents the extra eigenvalue sitting at $x=x_{1,n}$, while the
second term in \rhoform\ can be interpreted as the backreaction to
the Fermi sea in putting an eigenvalue at $x=x_{1,n}$
\HanadaIM.\foot{Recall also our interpretation below \naiveann\ of
the source term $J_{1,n}$ as encoding the effect of interactions
between the eigenvalue and the Fermi sea.}

Note that the coefficient of the $\delta$-function in \rhoform\ is
exactly one. It was emphasized in \HanadaIM\ that this is a
consequence of the Cardy condition. Another interesting feature of
\rhoform\ is that
\eqn\rhointzero{
\int_{-\infty}^\infty dx\,\rho_{1,n}(x)=0 ~.
}
This follows from the definition \rhomndef\ of $\rho_{1,n}(x)$ and
the fact that the minisuperspace amplitude $\Psi_{1,n}(\ell)$
vanishes at $\ell=0$. Of course, it can also be checked by the
direct integration of \rhoform. The relation \rhointzero\ suggests
the picture that exactly one eigenvalue is removed from the Fermi
sea $x\leq-1$ and placed at the point $x=x_{1,n}$ which is an
extremum of the effective potential.

For the $p=2$ models, the ZZ brane deformations are also extremely
simple. Substituting $p=2$ and \zonenpm\ into \FZZTZZdefsub, we
find
\eqn\ZZdefpeqii{
\delta_{1,n} F \sim {y\over (x-x_{1,n})\sqrt{x+1}} ~.
}
Remembering that the curve $y(x)$ takes the form
\eqn\curveptwo{
y^2=2^{q-2}(x+1)\prod_{l=1}^{{q-1\over2}}(x-x_{1,l})^2 ~,
}
we find
\eqn\delFxy{
\delta_{1,n} F \sim \prod_{l\not=n}(x-x_{1,l})~.
}
Note that the degree of the polynomial \delFxy\ is ${q-3\over2}$.
In particular, $\delta F(x,y)$ is a constant for the pure gravity
case $(p,q)=(2,3)$ \HanadaIM. From \delFxy, it is obvious that the
$(1,n)$ ZZ brane deformation vanishes at every singularity except
$x=x_{1,n}$; therefore it splits only the $(1,n)$ singularity.
Using \ZZdefpeqii, or equivalently
\eqn\ZZdefpeqiiy{
\delta_{1,n}y \sim {1\over (x-x_{1,n})\sqrt{x+1}} ~,
}
we can also make very concrete the discussion of the ZZ brane
$B$-cycle deformations at the end of section 3.1. In particular,
it is immediately clear from \ZZdefpeqiiy\ that the integral
$\delta_{1,n}\oint_{B_{1,n'}}y$ receives a logarithmic enhancement
for $n'=n$, since in this case the integral runs between $x\sim
-1$ and the branch point near the pole at $x= x_{1,n}$.

The ZZ brane deformations \delFxy\ have a nice interpretation in
the one-matrix model. Recall that at infinite $N$, before the
double-scaling limit, the matrix model curve can be written as
\eqn\loopeqn{
y^2= V^\prime(x)^2 +f(x)\;,
}
where $V(x)$ is the matrix model potential of minimal degree
${q+3\over2}$ and $f(x)$ is a polynomial of degree ${q-1\over2}$.
The main effect of the double-scaling limit is to scale away the
leading term in \loopeqn, reducing the degree of $y^2$ from $q+1$
to $q$. Now let us consider the relevant deformations of the curve
\loopeqn. We can either deform the potential or $f(x)$. Allowing
for shifts and rescalings of $x$, $V'(x)$ has ${q-1\over 2}$ free
parameters. These correspond to the ${q-1\over 2}$ tachyon
deformations. Since $y^2$ has $q-1$ free parameters modulo shifts
and rescalings of $x$, this leaves ${q-1\over2}$ deformations to
$f(x)$; these are of degree at most ${q-3\over 2}$. Comparing with
the degree of the ZZ brane deformations \delFxy, we conclude that
adding ZZ branes amounts to deforming $f(x)$ while keeping the
potential fixed. Since $f(x)$ determines how the eigenvalues are
distributed among the extrema of $V(x)$, this confirms that adding
ZZ branes corresponds to shifting eigenvalues from the Fermi sea
to the other extrema of the potential.

\vskip 0.5cm

\noindent {\bf Acknowledgments:}

We would like to thank J.~Gomis, A.~Hashimoto, S.Y.~Lee,
J.~Maldacena, E.~Martinec, G.~Moore, V.~Niarchos and D.~Sahakian
for useful discussions. The research of DK, KO and
 JP is supported in part by DOE grant DE-FG02-90ER40560.
 The research of NS is
supported in part by DOE grant DE-FG02-90ER40542. The research of DS is
supported in part by an NSF Graduate Research Fellowship and by NSF grant
PHY-0243680. Any opinions, findings, and conclusions or recommendations
expressed in this material are those of the author(s) and do not necessarily
reflect the views of the National Science Foundation.

\vfill

\appendix{A}{Other Annulus Amplitudes}

In this appendix, we will study the FZZT-FZZT annulus amplitudes
with more general matter states. These will provide a highly
non-trivial test of the FZZT identification formulas of
\SeibergNM,
\eqn\FZZTident{
|\sigma;k,l\rangle = \sum_{m'=-(k-1),2}^{k-1}\
\sum_{n'=-(l-1),2}^{l-1}|\sigma+{i(m'q+n'p)\over
\sqrt{pq}};1,1\rangle
}
that relate FZZT branes with general matter state to FZZT branes
with matter state $(1,1)$.

The annulus amplitude with general matter boundary conditions was
derived in \MartinecKA\ using the continuum Liouville approach:
\eqn\Zac{\eqalign{
Z(\sigma,a|\sigma',a')&= {1\over
p}\sum_{j=1}^{p-1}\int_{-\infty}^{\infty} d\nu {\sin (\pi a j
/p) \sin (\pi a' j / p)\cos (\pi\sigma\nu) \cos
(\pi\sigma'\nu) \over \nu
\sinh(\pi\nu/b)\Big(\cosh(\pi\nu/b)-\cos(\pi j/p)\Big)}
}}
where $a,a'(=1,\cdots, p-1)$ label the nodes (or ``heights'') of
the $A_{p-1}$ Dynkin diagram \PasquierJC. Note that the matter
boundary condition with height $a$ corresponds to the minimal
model Cardy state $|k,l\rangle =|a,1\rangle$. We wish to simplify
the annulus amplitudes \Zac\ as in section 2.1. First, we evaluate
the sum over $j$ in \Zac\ using the identity\foot{To prove this
identity, notice that both sides are meromorphic functions of
$\nu$ with the same poles and residues. Therefore they differ by a
holomorphic function $f(\nu)$. Since both the LHS and the RHS tend
to zero as $\nu\to+\infty$ away from the imaginary $\nu$ axis,
$f(\nu)$ must be identically zero.}
\eqn\sumj{
{1\over p}\sum_{j=1}^{p-1} {\sin (\pi a j /p) \sin (\pi a' j /
p)\over \cosh(\pi\nu/b)-\cos(\pi j/p)} = {\sinh(\pi
a\nu/b)\sinh(\pi (p-a')\nu/b)\over \sinh(\pi\nu/b)\sinh(\pi
p\nu/b)},\qquad (a,a'\in \Bbb Z^+,\ a\le a')~.
}
Then the annulus amplitude \Zac\ becomes
\eqn\Zacbec{
Z(\sigma,a|\sigma',a') =\int_{-\infty}^{\infty} d\nu {\cos
(\pi\sigma\nu) \cos (\pi\sigma'\nu) \sinh(\pi a\nu/b)\sinh(\pi
(p-a')\nu/b) \over \nu\sinh^2(\pi\nu/b)\sinh(\pi
p\nu/b)},\qquad(a\leq a')~.
}
The integral can be evaluated by regularizing
\eqn\regularize{
{1\over \nu} \to {\nu\over \nu^2+\varepsilon^2}
 }
and closing the contour of $\nu$-integral as in section 2.1. For
$a,a'=1,p-1$ this yields (up to an additive constant)
 \eqn\ZacII{\eqalign{
&Z(\sigma,1|\sigma',1)=Z(\sigma,p-1|\sigma',p-1)
=\log\left({z-z'\over x-x'}\right) ~,\cr
 &Z(\sigma,1|\sigma',p-1)=-\log(z+z')~. }}
In general, we find a somewhat more complicated formula
\eqn\ZacIII{\eqalign{
&Z(\sigma,a|\sigma',a') = -\sum_{k=1}^{\infty} {2\over
k}U_{a-1}(\cos{\pi k\over p})U_{a'-1}(\cos{\pi k\over p}) e^{-{\pi
k\sigma\over\sqrt{pq}}} \cosh {\pi k\sigma'\over\sqrt{pq}} \cr
    & + a\sum_{k=1}^{\infty}
 {2\over  k} (-1)^{(a+a')k}e^{-{\pi p k\sigma\over\sqrt{pq}}} \cosh {\pi
 p k\sigma'\over\sqrt{pq}} +\left({a(p-a')\over
 \varepsilon\sqrt{pq}}-a(p-a'){\pi\sigma\over\sqrt{pq}}\right)\cr
 &\qquad\qquad\qquad\qquad\qquad\qquad\qquad\qquad\qquad (\sigma>\sigma'>0,\ a\le a') ~.
}}

Now let us compare \ZacIII\ with the result of using the FZZT
identifications \FZZTident. We define
\eqn\Zacident{\eqalign{
\tilde Z(\sigma,a|\sigma',a')&\equiv \sum_{m=-(a-1),2}^{a-1}\
\sum_{m'=-(a'-1),2}^{a'-1}Z(\sigma+{im\over b},1|\sigma'+{im'\over
b},1) ~.
}}
Then one can show that
\eqn\Zacdiff{\eqalign{
Z(\sigma,a|\sigma',a')-\tilde Z(\sigma,a|\sigma',a')
    &=-{a(a'-1)p\over\varepsilon\sqrt{pq}}+a(a'-1)p{\pi\sigma\over\sqrt{pq}}\cr
    &\qquad -a(a'-1)\sum_{k=1}^{\infty}{2\over k}(-1)^{(a+a')k}e^{-{\pi p k\sigma\over\sqrt{pq}}} \cosh {\pi
 p k\sigma'\over\sqrt{pq}}\cr
 &=-{a(a'-1)p\over\varepsilon\sqrt{pq}}+a(a'-1)\log\big( 2(x-(-1)^{a+a'}x')\big)~. \cr
 &\qquad\qquad\qquad\qquad\qquad\qquad\qquad\qquad\qquad (a\le a')
}}
In other words, the actual annulus amplitude \ZacIII\ differs from
the result of applying the FZZT identifications by a sum over the
unphysical states with $k=0\,\mod\,p$. According to the discussion
at the end of section 2.1, we expect this discrepancy to be
non-universal and analytic in $\mu$. Indeed, if we restore the
powers of $\mu$ by writing $x=\mu_B/\sqrt{\mu}$ and similarly for
$x'$, and in addition make the correspondence\foot{This
correspondence is motivated by the fact that the divergence as
$\varepsilon\to0$ is precisely an infinite volume divergence, while
${1\over 2b}\log\left({\Lambda\over\mu}\right)$ is the usual
expression for the ``volume'' of the Liouville direction cut off by
the Liouville wall at $\phi\to+\infty$.}
\eqn\epsid{
{1\over\varepsilon} \to {1\over
2b}\log\left({\Lambda\over\mu}\right) ~,
}
then \Zacdiff\ becomes
\eqn\ZacdiffII{\eqalign{
Z(\sigma,a|\sigma',a')-\tilde Z(\sigma,a|\sigma',a')
    &=a(a'-1)\log\big(
    2(\mu_B-(-1)^{a+a'}\mu_B')\big)-{1\over2}a(a'-1)\log{\Lambda}
 }}
i.e.\ the $\mu$ dependence drops out all together. Therefore, we
have found that the FZZT identifications predict the correct
annulus amplitudes up to terms analytic in $\mu$. Since these
analytic terms are regularization dependent, the FZZT
identifications in some sense provide an alternate regularization
of the annulus amplitude.

\appendix{B}{Normalization of the ZZ boundary states}

In this appendix, we will discuss the proper normalization of the
ZZ boundary states in minimal string theory. We will be particularly
interested in the relative normalization of the ZZ and FZZT branes.
In Liouville CFT, this normalization is fixed to be \ZamolodchikovAH
\eqn\ZZnormL{
|m,n\rangle_{\rm Liouville} = |\sigma_{m,n}\rangle_{\rm
Liouville}-|\sigma_{m,-n}\rangle_{\rm Liouville}
 }
by the requirement that the ZZ-FZZT annulus amplitude must be
given by a trace over open-string states stretched between the two
branes, with every state contributing $+1$ to the partition
function. As we will see below, in minimal string theory the
relative normalization can be fixed by target space
considerations, and it is different from \ZZnormL.

Consider first the natural normalization of the FZZT boundary
state in Liouville theory. The one point function of the cosmological
constant on the disk in this state, with a $(1,1)$ boundary state for
matter, is given by\foot{We use the normalization given in
\ZamolodchikovAH; it differs by a minus sign from \FateevIK.}
\refs{\FateevIK,\ZamolodchikovAH,\AlexandrovNN}:
\eqn\opfmu{\eqalign{
\left\langle e^{2b\phi}(0)\right\rangle
 &= -A_M{2^{1\over4}\over \sqrt\pi}
\sqrt{q\over p}\ {\Gamma({p\over q})\over \Gamma({q\over
 p})}{1\over\sin{q\pi \over p}}(\sqrt{\mu})^{q/p-1}\cosh
 \left(b-{1\over b}\right)\pi\sigma~.\cr
 }}
$A_M$ is the contribution to the one-point function from the
matter sector (we assume the $(1,1)$ matter state): \eqn\AMSmat{
A_M=\sqrt{|S_{(1,1),(1,1)}|}=\left({8\over pq}\right)^{1\over4}
\left|\sin{\pi(q-p)\over p}\sin{\pi (q-p)\over q}\right|^{1\over
2}. } Here $S_{(1,1),(1,1)}$ is an element of the modular $S$
matrix \Smat. It is always positive for the unitary models
($q=p+1$), so in these cases the absolute value in \AMSmat\ can be
omitted. For some non-unitary theories, $S_{(1,1),(1,1)}$ is
negative; in these cases the absolute value in \AMSmat\ ensures
that the one point function \opfmu\ is real. We will see
momentarily that this choice of the phase of the boundary state is
natural from the target space point of view as well.

Integrating \opfmu\ with respect to $-\mu$ holding
$\mu_B=\sqrt{\mu}\cosh\pi b\sigma$ fixed, and remembering
the rescaling of $\mu$ by $\pi\gamma(b^2)$, we find the
FZZT disk amplitude:
\eqn\fzztdisk{
Z(\sigma) = -A_M{2^{5\over4}\over \pi^{3\over2}} {\sqrt{pq}\over
p+q}{\Gamma(1-{p\over q})\over \Gamma({q\over
p})}{1\over\sin{q\pi\over p}}(\sqrt{\mu})^{q/p+1}\Big( \sinh  \pi
b\sigma \sinh {\pi \sigma\over b}-b^2 \cosh\pi b\sigma \cosh {\pi
\sigma\over b}\Big)
 }
We can now check that the choice of phase of the boundary state
implicit in \fzztdisk\ is the one natural from the target space
point of view. As discussed in \SeibergNM, the FZZT brane can be
identified with the effective potential of a probe eigenvalue in
the double scaling limit $Z(\sigma)=-{1\over 2} V_{\rm eff}(x)$.
Thus, for example, we expect it to be unbounded from below for
unitary theories. It is easy to check that for $q=p+1$, \fzztdisk\
indeed has the property that as $\sigma\to\infty$ (or equivalently
$x\to\infty$ \introsigma),
 \eqn\behsig{Z(\sigma)\sim C\exp(\pi\sigma(b+{1\over b}))}
with $C$ a positive constant. Hence the effective potential
$V_{\rm eff}\to-\infty$ in this limit, in agreement with the
original construction of the double scaling limit
\refs{\DouglasVE\GrossVS\BrezinRB-\DouglasDD}.

As another check on the phase of \fzztdisk, consider the
case $(p,q)=(2,2k-1)$. In this case, one again finds the
behavior \behsig, with $C_k\sim (-1)^k$. Thus, the effective
potential is not bounded from below for even $k$ and goes
to $+\infty$ as $x\to\infty$ for odd $k$. This, again, is
consistent with the known results from double scaled matrix
models. Models with $k$ even (such as pure gravity, corresponding
to the case $k=2$) correspond to matrix integrals with potentials
that are not bounded from below, whereas for $k$ odd the
potential is bounded from below and the matrix integral is well
defined.

After understanding the natural phase of the FZZT boundary state,
we move on to the ZZ branes. These branes give non-perturbative
instanton corrections to string amplitudes. One can check that
this implies the ZZ boundary states should be defined with an
extra minus sign relative to the CFT definition \ZZnormL:
\eqn\ZZnormfin{ |m,n\rangle = |\sigma_{m,-n}\rangle -
|\sigma_{m,n}\rangle }
As an example, consider again the unitary case $q=p+1$. Using
the modified prescription \ZZnormfin, one finds
\eqn\Zmnunitary{\eqalign{
Z_{m,n} &= Z(\sigma_{m,-n}) -Z(\sigma_{m,n})\cr
 &= -A_M{2^{9\over4}\over \pi^{3\over2}}
{\sqrt{pq}\over p+q}{\Gamma(1-{p\over q})\over \Gamma({q\over
p})}{1\over\sin{\pi\over p}}(\sqrt{\mu})^{q/p+1}\sin
 {m\pi \over p}\sin{n\pi \over p+1} <0 ~.
}}
As indicated in \Zmnunitary, all $Z_{m,n}$ are negative in this
case, in agreement with the fact that the leading non-perturbative
corrections, which go like $\exp(Z_{m,n})$, give rise to
$\exp(-1/g_s)$ instanton effects.

For non-unitary models, \ZZnormfin\ implies that some of the ZZ
branes have in general positive $Z_{nm}$ or negative $V_{\rm
eff}$. By the usual rules of D instantons this implies that in
these cases there are non-perturbative effects that go like
$e^{+\CO(1/g_s)}$ instead of $e^{-\CO(1/g_s)}$. The existence of
such catastrophic non-perturbative effects seems to be related to
the fact that in non-unitary theories the perturbative vacuum is
an unstable critical point of the effective potential $V_{\rm
eff}$, and there are nearby lower critical points. For example,
when $p=2$, one can show \SeibergNM\ that $V_{eff}$ has a number
of local minima (corresponding to the $(1,n)$ ZZ branes with $n$
odd) with energies {\it below} the Fermi level. Therefore the
perturbative vacuum is unstable to eigenvalues tunneling to these
local minima. It is important to emphasize that this phenomenon is
separate from the question of whether the matrix model potential
is bounded from below at infinity.

One interesting consequence of \ZZnormfin\ is that there is an
extra minus sign in the ZZ-FZZT annulus, relative to Liouville
theory alone. This means that the open strings stretched between
ZZ and FZZT branes are {\it fermions} in minimal string theory! It
is, of course, very surprising to find fermions emerging out of
bosonic string theory. But this interpretation is well-supported
by the matrix model, where the FZZT branes are described by the
determinant operator $\det(x-M)$. We can also write this
determinant as a Grassman integral over $N$ complex fermions
$\psi_i$ \eqn\detrew{ \det(x-M) = \int d\psi^\dagger d\psi\
e^{\psi^\dagger(x-M)\psi} } In the one-matrix model, we interpret
the matrix $M$ as describing the (bosonic) open strings stretched
between the $N$ ZZ branes in the Fermi sea. Therefore, the
$\psi_i$ represent the fermionic open strings stretched between
the FZZT brane and the $N$ ZZ branes. They are complex because of
the two different orientations of these open strings. The
parameter $x$ which labels the FZZT brane and corresponds to the
boundary cosmological constant appears here as a mass term for the
fermions.  It can be thought of as representing the length of the
corresponding open strings.

\listrefs

\end